\documentclass[10pt,journal,compsoc]{IEEEtran}
\usepackage{balance}  

\usepackage{setspace}

\usepackage{amsmath}
\usepackage{amsfonts}
\usepackage{amssymb}
\usepackage{amsthm}
\usepackage{balance}
\usepackage{url}

\let\emptyset\varnothing

\usepackage[ruled,vlined]{algorithm2e}
\renewcommand{\CommentSty}[1]{\textnormal{#1}}
\DontPrintSemicolon
\SetKwComment{tcp}{$\triangleright$ }{}
\SetVlineSkip{0cm}
\SetKwInOut{Output}{Output}
\SetKwInOut{Input}{Input}

\usepackage{graphicx}

\usepackage{subfigure}
\usepackage{wrapfig}

\theoremstyle{definition}
\newtheorem{definition}{Definition}

\usepackage{color}

\usepackage{colortbl}
\usepackage{multicol}
\usepackage{multirow}
\renewcommand{\mid}{:}

\usepackage{footnote}
\makesavenoteenv{tabular}
\makesavenoteenv{table}

\usepackage{xargs}
\usepackage{xspace}
\usepackage[pdftex,dvipsnames]{xcolor}  
\usepackage[colorinlistoftodos,prependcaption,textsize=tiny]{todonotes}

\newcommandx{\uvc}[2][1=]{\todo[color=magenta!50,#1]{\sf \textbf{\"Umit:} #2}\xspace}

\newcommandx{\ay}[2][1=]{\todo[color=blue!50,#1]{\sf \textbf{Apo:} #2}\xspace}

\newcommandx{\sr}[2][1=]{\todo[color=blue!50,#1]{\sf \textbf{Siva:} #2}\xspace}

\definecolor{darkgreen}{rgb}{0,0.5,0}
\definecolor{midnight}{rgb}{0,0.094,0.533}
\definecolor{ocean}{rgb}{0,0.290,0.533}

\usepackage[colorlinks=true,%
   linkcolor=red,%
   citecolor=midnight,%
   urlcolor=blue,%
   bookmarks=false]{hyperref}

\usepackage[sort,compress]{cite}

\usepackage{pifont}
\newcommand{\cmark}{\text{\ding{51}}}
\newcommand{\xmark}{\text{\ding{55}}}

\definecolor{darkgreen}{rgb}{0,0.4,0}

\usepackage{tikz}
\usetikzlibrary{decorations.pathreplacing,calc}

\newcommand\smallfont{\fontsize{10}{11}\selectfont}

\newcommandx{\pgabb}{\ensuremath{\text {\sc PGAbB}}\xspace}

\newcommandx{\bbtc}{\ensuremath{\text {\sc bbTC}}\xspace}

\newcommandx{\orderdeg}{\ensuremath{\text {\sc orderByDeg}}\xspace}
\newcommandx{\intersect}{\ensuremath{\text {\sc nIntersect}}\xspace}
\newcommandx{\tci}{\ensuremath{\text {\sc TC-Intersect}}\xspace}
\newcommandx{\tch}{\ensuremath{\text {\sc TC-HMap}}\xspace}
\newcommandx{\bbtci}{\ensuremath{\text {\sc bbTC-Intersect}}\xspace}
\newcommandx{\bbtch}{\ensuremath{\text {\sc bbTC-HMap}}\xspace}
\newcommandx{\bbtask}{\ensuremath{\text {\sc bbTaskComposition}}\xspace}
\newcommandx{\tgpu}{\ensuremath{\text {\sc runTaskOnGPU}}\xspace}
\newcommandx{\tcpu}{\ensuremath{\text {\sc runTaskOnCPU}}\xspace}

\begin{document}

\title{A Block-Based Triangle Counting Algorithm on Heterogeneous Environments}

\author{
    Abdurrahman Ya\c{s}ar,
    Sivasankaran Rajamanickam~\IEEEmembership{(Member,~IEEE)},
    Jonathan Berry, and\\
    \"{U}mit V. \c{C}ataly\"{u}rek~\IEEEmembership{(Fellow,~IEEE)}

\IEEEcompsocitemizethanks{\IEEEcompsocthanksitem Ya\c{s}ar, and \c{C}ataly\"{u}rek
are with the School of Computational Science and Engineering at Georgia Institute
of Technology.\protect\\
E-mail: \{ayasar,umit\}@gatech.edu
\IEEEcompsocthanksitem Rajamanickam, and Berry are with
the Center for Computing Research at Sandia National Laboratories, Albuquerque,
NM.\protect\\
E-mail: \{srajama,jberry\}@sandia.gov
}

\thanks{Manuscript received xx xx, xx; revised xx xx, xx.}}

\IEEEtitleabstractindextext{%
\begin{abstract}
Triangle counting is a fundamental building block in graph algorithms.
In this paper, we propose a block-based
triangle counting algorithm to reduce data movement during both sequential and
parallel execution.
Our block-based formulation makes the
algorithm naturally suitable for heterogeneous architectures.
The problem of partitioning the adjacency matrix of a graph is well-studied.
Our task decomposition goes one step further: it partitions the set of
triangles in the graph.  By streaming these small tasks to compute resources,
we can solve problems that do not fit on a device.
We demonstrate the effectiveness of our approach by providing an implementation
on a compute node with multiple sockets, cores and GPUs.
The current state-of-the-art in triangle enumeration processes the
Friendster graph in 2.1 seconds, not including data copy time between
CPU and GPU. Using that metric, our approach is 20 percent faster.
When copy times are included, our algorithm takes 3.2 seconds.
This is 5.6 times faster than the fastest published CPU-only time.
\end{abstract}

}

\maketitle

\IEEEdisplaynontitleabstractindextext
\IEEEpeerreviewmaketitle

\section{Introduction}
\label{sec:intro}

Graphs are very useful data-structures. They can represent different
applications, such as social network analytics, biological networks, and scientific
simulations. Today there exist large graphs that have billions of vertices and
edges. High-performance processing of these large graphs is crucial and pervasive.
Most of the current high-performance solutions rely on distributed systems because
one can process problem instances that are bigger than the main memory of a single
node.
However, communication and data replication in distributed systems often leads
to bottlenecks.  The popularity of multi-core machines
increased drastically during the past decade.  Today, different hardware vendors
have developed processors with multiple cores to deliver improved
performance, and hence the multi-core technology became ubiquitous.
In addition to multi-core technology, many hardware accelerators like graphics
processing units (GPUs), and field-programmable gate arrays (FPGAs) also became a part
of the systems to serve different parallelization needs, and more are coming.
Soon, we will have distributed systems consisting of multi-core servers with many
different types of hardware accelerators. Such an environment increases the
importance of designing flexible algorithms for performance-critical kernels and
implementations that can run well on various platforms. In this work, we propose
an architecture-agnostic triangle counting algorithm, \bbtc, and we do an
extensive analysis of its performance on a single node multi-CPU and multi-GPU
execution environment.

\begin{figure}[ht]
\centering
  \subfigure[A triangle]{\includegraphics[width=.25\linewidth]{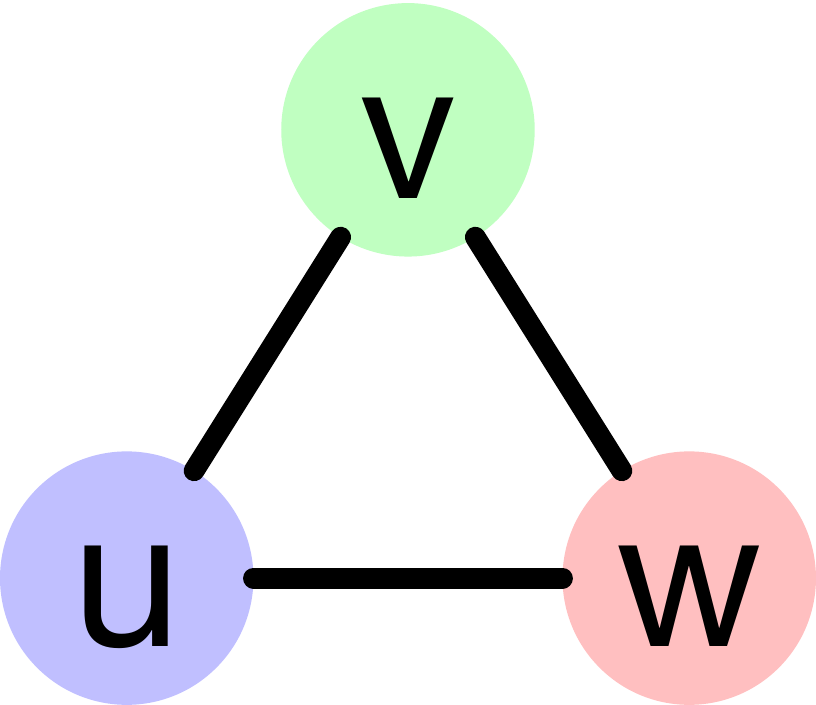}
  \label{fig:ex-tri}}

  \subfigure[A 5-way 1D partition]{\includegraphics[width=.45\linewidth]{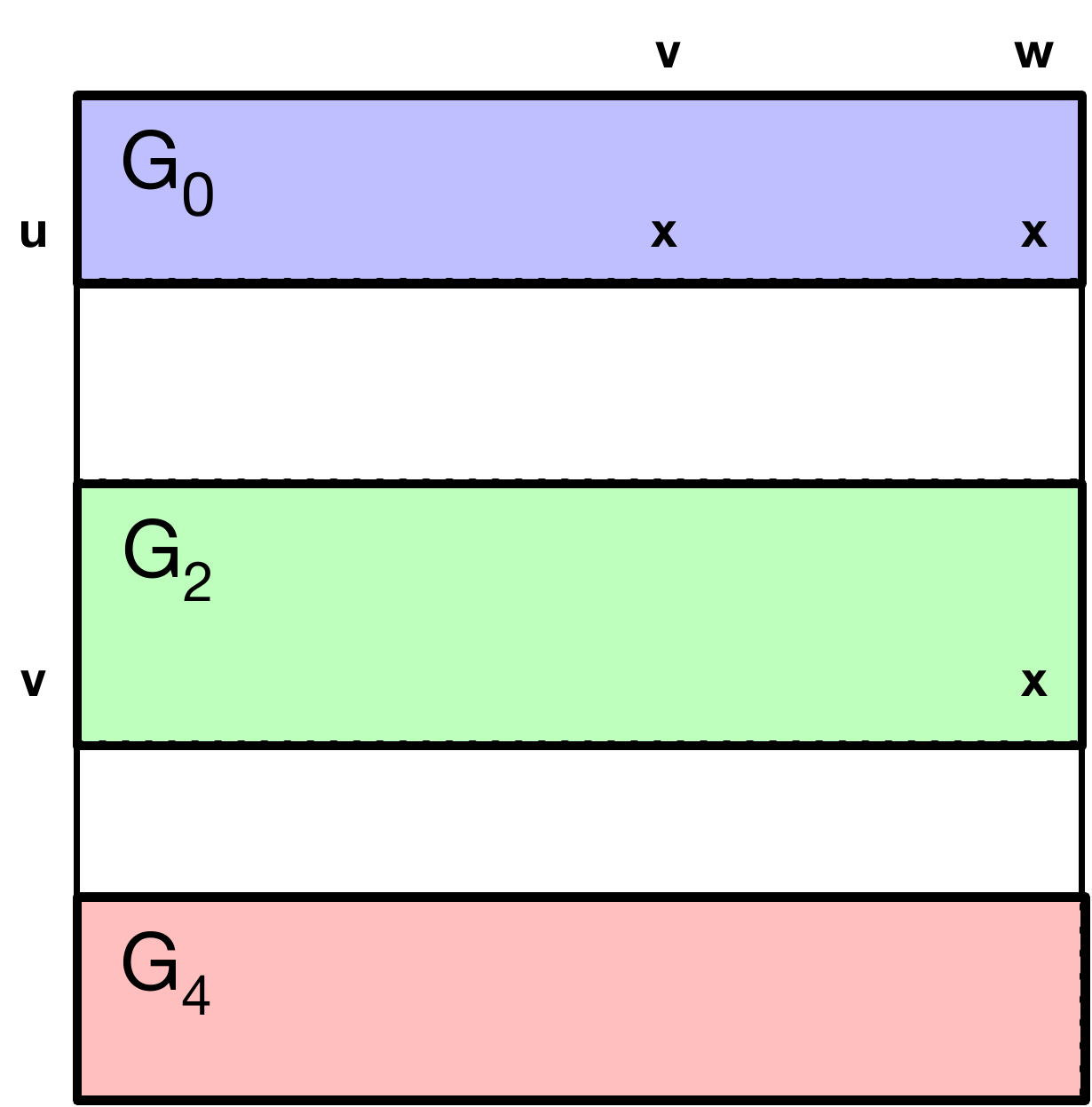}
  \label{fig:ex-tri-1D}}
  \hspace{1ex}
  \subfigure[A 5x5-way 2D partition]{\includegraphics[width=.45\linewidth]{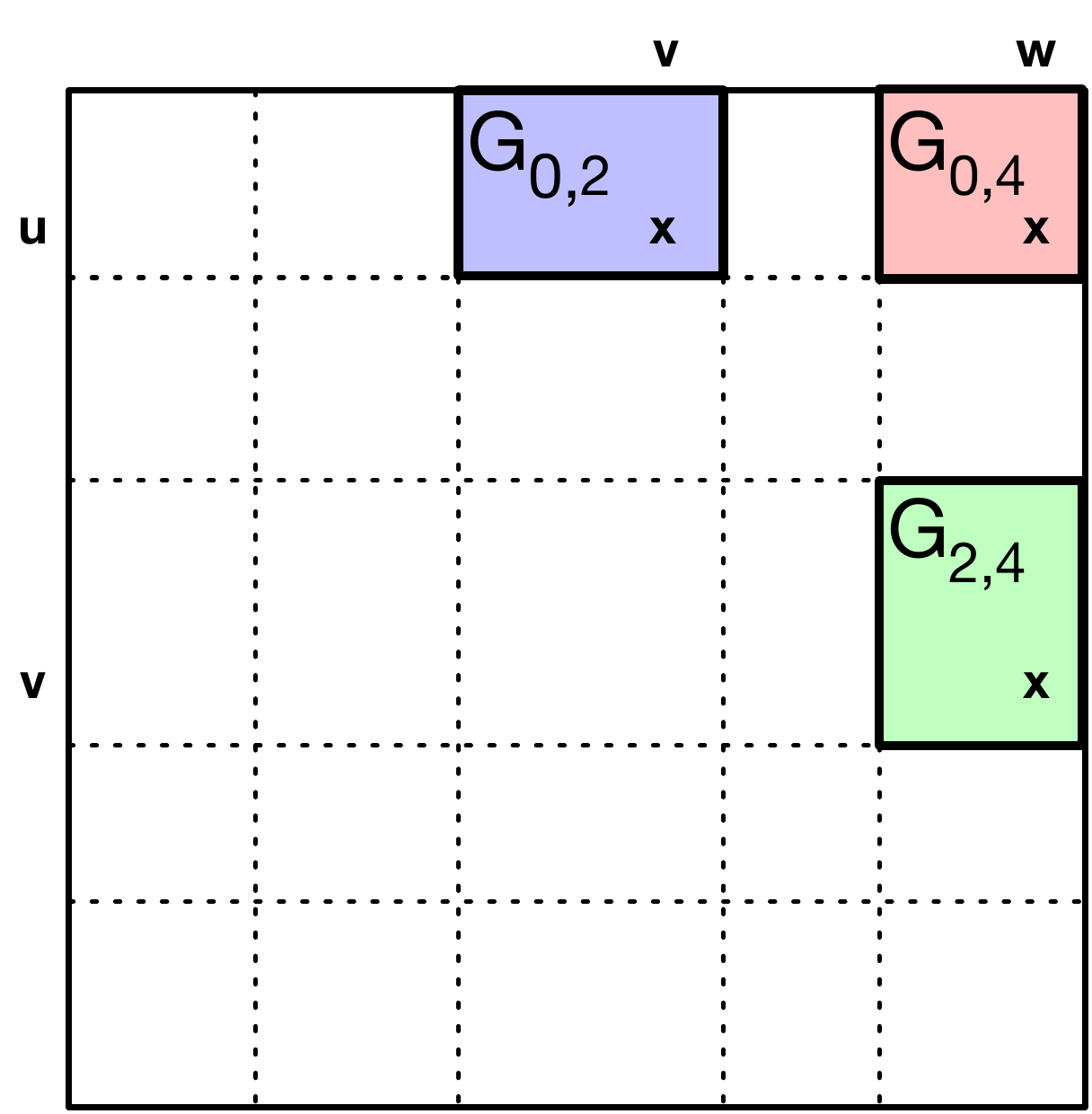}
  \label{fig:ex-tri-2D}}

  \caption{Finding graph triangles in 1D and 2D-partitioned adjacency matrices.}
  \label{fig:tri-example}
\end{figure}

The triangle counting problem~\cite{Latapy08-TCS,Shun15-ICDE,Hu18-SC,Dhulipala18-SPAA,Berry14-TCS,Itai78-SJC,Ortmann14-ALENEX,Alon97-Algorithmica,Pagh14-PODS}
is to find the number of 3-cliques in an undirected graph (Fig.~\ref{fig:ex-tri}).
This is a crucial graph kernel that serves as a building block for many other
graph problems such as k-truss decomposition~\cite{Cohen08-NSATR}, community
detection~\cite{Prat12-CIKM}, thematic structure identification~\cite{Eckmann02-NAS},
dense sub-graph discovery~\cite{Wang10-VLDBJ} and link recommendation~\cite{Tsourakakis11-SNAN}.
In recent years, algorithmic and architectural advances have led to great
improvements in triangle counting performance~\cite{Date17-HPEC,Yasar19-HPEC,Hu18-SC,Green14-WIAP}.
None of these approaches are truly heterogeneous because they all target one
architectural feature. In Section~\ref{sec:bbtc}, we propose a block-based
approach for heterogeneous execution environments that simultaneously leverages
both CPUs and multiple GPUs. Such an approach is crucial for achieving three
goals: maximizing memory utilization, processing graphs that cannot fit into
GPU memory and overlapping the data-copy time with the computation.
To achieve these goals, we tackle the triangle counting problem under three
axes; data/computation partitioning, a block-based formulation for the triangle
counting problem, and architecture agnostic task execution.

At the abstract level, our execution model is as follows. Computation is divided
into tasks. Each task depends on multiple blocks. A light-weight scheduler
schedules tasks on available devices. Scheduler is also responsible for the
movement of data blocks that are needed for execution of each individual task.

Execution time is a function of the computational load of individual tasks, and
communication costs. The granularity of the task decomposition is crucial for
performance.
Therefore, partitioning is the first step and refers to both data and
computation partitioning.
In the context of graphs, we can do the partitioning at the vertex level (coarse-grained) or
the edge level (fine-grained). A hybrid approach may also be used.
There is a duality between graphs and sparse matrices. Vertex-level and edge-level
partitionings correspond to row (1D) and
nonzero (2D) partitionings. For visualization
purposes, consider the adjacency matrix representation of a graph.
Fig.~\ref{fig:ex-tri-1D} illustrates a 1D partitioning. This generates
coarse-grained tasks. Some state-of-the-art
CPU-based triangle counting algorithms (such as~\cite{Yasar18-HPEC,Shun15-ICDE})
generate that kind of coarse-grained tasks. However, such a task generation
ends up with higher copy/communication costs on a heterogeneous
environment and highly imbalanced workload
distribution~\cite{Boman13-SC,Saule12-JPDC-spart,Catalyurek10-SISC,kayaaslan20181,acer2018optimizing}.
We can address this problem by using a 2D partitioning strategy and generating
medium-grained tasks (e.g., see Fig.~\ref{fig:ex-tri-2D}) to solve the triangle counting problem.
Generally, there are two classes of algorithms that provide such
partitioning: i)~connectivity-based techniques (graph and hypergraph)~\cite{Catalyurek10-SISC},
ii)~spatial partitioning techniques~\cite{Saule12-JPDC-spart}.
Connectivity-based  techniques find a reordering of the vertices
while spatial techniques break nonzeros of the matrix into blocks without
reordering the vertices.

We seek partitioning methods that maximize memory efficiency
and minimize data movement
which is essential to overlap the communication with the computation.
Unfortunately, unstructured or random partitionings may
easily end up with highly irregular data access patterns (i.e., one block may
require to access all other blocks). Our ultimate goal is to localize sub-graph
computations, by improving data access patterns and data movement in the system
architecture. Therefore we need a special partitioning scheme that we could use for
divide-and-conquer approaches. For this reason, we utilize a special 2D partitioning
scheme called {\em symmetric rectilinear partitioning} (also known as
{\em symmetric generalized block distribution} (see Fig.~\ref{fig:ex-tri-1D}).
This problem, first defined by Grigni and Manne~\cite{Grigni96-PAISP}. Recently,
Ya\c{s}ar and \c{C}ataly\"{u}rek~\cite{Yasar19-arXiv} proposed some heuristic
algorithms to solve this problem.
It is also possible to achieve such partitioning by utilizing more complex
connectivity-based~\cite{Catalyurek10-SISC} techniques. However, these techniques have two drawbacks; first,
most of these techniques are computationally expensive, second, connectivity-based
ordering of the vertices may affect the computational complexity of the algorithm.
In sparse graphs with heavy-tailed degree distributions, triangle counting
has low computational complexity~\cite{Berry14-TCS}
therefore we need
a fast partitioning technique. In this work, we use a parallel version of the
PBD algorithm presented in~\cite{Yasar19-arXiv} to partition the graph into blocks.

We propose a medium-grained triangle counting formulation (\bbtc) on top of
symmetric rectilinear partitioning. This makes the algorithm naturally suitable for
task-based execution on shared and distributed-memory systems as well as on
heterogeneous architectures.
\bbtc defines tasks based on generated blocks. CPUs and GPUs
process tasks together.
Although the creation of tasks is dependent on graph data,
our algorithm to dispatch and execute these tasks is data-agnostic.
This agnostic structure will allow execution on heterogeneous settings with any
accelerator and CPU combination.

One can further optimize the execution time by
utilizing different data structures and implementation of the graph kernel based
on properties of the input data and architectures. However, in this work, we
utilize a single data-structure and just one implementation of each task type on
each processor type. The power of the algorithm comes from partitioning of the
triangle space into tasks, a dynamic task-scheduling
scheme that schedules tasks on multi-core-CPUs, use of multiple streams on multiple
GPUs to effectively utilize the massive computing capabilities on the GPUs, and, at
the same time, move the data to the GPUs asynchronously.

The primary contributions of this work can be listed as:

\begin{itemize}
\item We show how to partition a graph's triangle set, and we
use this observation to derive a new block-based triangle counting
algorithm.

\item We show how to run our algorithm on heterogeneous hardware when
the graph does not fit on device.

\item We propose dynamic scheduling techniques for block-based execution model
for heterogeneous environments and show how tasks can be distributed among
different CPUs and GPUs efficiently.

\item We evaluate characteristics of our algorithm in theoretical and
practical aspects and perform extensive analysis on different architectures.
\end{itemize}

The preliminary results of this work were presented as an extended abstract
in~\cite{Yasar19-HPEC}.
This paper presents a detailed description of our algorithm and its complexity
analysis, a thorough experimental evaluation of the components of the algorithm and
comparisons against the state-of-the-art. These results demonstrate that our
implementation beats the fastest published end-to-end CPU execution times.
The performance improvement is up to $11 \times$ over our
previous state-of-the-art implementation~\cite{Yasar18-HPEC}, $16 \times$ over
another state-of-the-art implementation~\cite{Shun15-ICDE}, and $1.5 \times$ over
state-of-the-art multi-GPU implementation~\cite{Hu18-SC} on the largest three
graphs.

\section{Problem Formulation}
\label{sec:formulation}

An undirected graph $G=(V, E)$, consists of a set of vertices $V$ and a set of
edges $E$. An edge $e$ is referred to as $e = (u, v) \in E$, where $u, v \in V$.
In the context of triangle counting, efficient algorithms traverse vertices
and edges following a complete ordering, which avoids double counting and also
reduces the number of edges traversed. For the sake of simplicity, in this work, we will
use vertex IDs for that complete ordering in the traversal.
Since G is undirected, we represent each edge as $(u,v)$, where $u < v$.
Under this representation, we call $u$ the \emph{source} vertex and $v$
the \emph{destination} vertex.  Thus, we induce an ordering that is
helpful when explaining our partitioning.
The neighbor list of a
vertex $u \in V$ is also defined using the same complete ordering, i.e.,
$N(G, u) = \{v\in V\mid(u, v)\in E$ and $u<v \}$. The degree of a vertex
$u \in V$ is defined as $d(G, u) = |N(G,u)|$. We will use $n$ and $m$ for number
of vertices and edges, respectively, i.e., $n=|V|$ and $m=|E|$.

\begin{definition}{Triangle Counting Problem.}
Given an undirected graph $G=(V, E)$, the triangle counting problem computes the
number of unique three cliques, i.e., the number of mutually connected three
vertices, in the graph, $G$.
\end{definition}

Given a graph $G$, and an integer $p$ such that $1 \leq p \leq n$. Let $V^{p}$
be a vertex partition set of $V$ into $p$ non-overlapping partitions
such that; $V^{p}=\{ V_0, \dots, V_{p-1}\}$ where $\cup_{0\leq i \leq p-1} V_i = V$
and $\cap_{0\leq i \leq p-1} V_i = \emptyset$.

Let $G_{i,j}$ be a sub-graph of $G$ such that
$G_{i,j} = \{(u,v) \in E \mid u \in V_i \wedge v \in V_j\}$.
By definition $G = \cup_{0\leq i,j\leq p-1} G_{i,j}$.
Our $G_{i,j}$ construct allows us to avoid double-counting any triangles.
In the partitioned graph we will only use subgraphs ($G_{i,j}$) that appear
in upper triangular part of the adjacency matrix, ($i \leq j$)
which allow us to only traverse edges $(u, v) \in G_{i,j}$, where $u < v$ by definition.
Source and destination vertex sets in a given sub-graph $G_{i,j}$,
are defined as $V_s(G_{i,j})=V_{i}$ and $V_d(G_{i,j})=V_{j}$, respectively.
The partial neighbor list of a vertex $u \in G_{i,j}$ is defined
as $N(G_{i,j}, u)$.
The partial degree of a vertex $u \in V_s(G_{i,j})$ is defined
as $d(G_{i,j},u) = |N(G_{i,j}, u)|$.
All the notations are summarized in Table~\ref{tab:notation}.

\begin{table}[bhpt]
  \caption{Notation used in this paper.}
  \label{tab:notation}
  \begin{small}
  \begin{tabular}{r  l}
  \textbf{Symbol} & \textbf{Description}                \\
  \hline \\
  $G=(V, E)$      & A graph $G$ with vertex \\
                  & \ and edge sets, $V$ and $E$, respectively \\
  $n=|V|$         & Number of vertices \\
  $m=|E|$         & Number of edges \\
  $V^{p}$         & Vertex partition set; $V^{p} = \{V_1, \dots, V_p\}$ \\
  $V_s(G_{i,j})=V_i$  & Source vertices of the subgraph $G_{i,j}$\\
  $V_d(G_{i,j})=V_j$  & Destination vertices of the subgraph $G_{i,j}$\\
  $N(G,u)$        & Neighbor list of vertex $u\in V$\\
  $N(G_{i,j},u)$  & Partial neighbor list of vertex $u$, i.e., \\
                  & $N(G_{i,j},u) = \{ (u,v) \mid  u \in V_i \wedge v \in V_j\}$\\
  $d(G,u)$        & Degree of vertex $u$, $d(G,u)=|N(G,u)|$\\
  $d(G_{i,j},u)$  & Partial degree of vertex $u \in V_s(G_{i,j})$, \\
                  & $d(G_{i,j},u)=|N(G_{i,j},u)|$ \\
  $t$             & A task; $t= \{i,j,k\} = \{G_{i,j}, G_{j,k}, G_{i,k}\}$\\
                  &  where $i\leq j \leq k$\\
  $T = \{t_0, \dots, t_i\}$  & Set of all tasks\\
  \hline
  \end{tabular}
  \end{small}
  \end{table}

\section{Work Optimal Triangle Counting Algorithms}

Sequential triangle counting is a well-studied
problem~\cite{Latapy08-TCS, Berry14-TCS}. The best algorithms rely on matrix
multiplication and run in $O(n^\omega)$ or $O(m^{\omega / (1+\omega)})$ where
$\omega$ is the matrix multiplication
exponent~\cite{Itai78-SJC,Alon97-Algorithmica}. However, these algorithms are
impractical and require $\theta(n^2)$ space. The fastest algorithm that doesn't
require $\theta(n^2)$ space does $O(m^{3/2})$ work~\cite{Latapy08-TCS}. Merge-based
and hashmap-based versions can both result in work optimal algorithms.
Both algorithms process vertices based on their degrees to bound the maximum
degree in the graph and guarantee $O(m^{3/2})$ work. This bound can be
improved if the degree distribution is governed by a power law with exponent
$\alpha$. In such a case, Latapy~\cite{Latapy08-TCS} shows that triangle
enumeration is bounded by $\Theta(mn^\frac{1}{\alpha})$. This is
an asymptotic improvement over the worst-case. Berry, et al.~\cite{Berry14-TCS}
improved upon this further and show that complexity of this problem can
actually be $\Theta(n)$ in realistic circumstances where the $4/3$ moment of
a graph is bounded by a constant.

\begin{algorithm}[t]
  \tcp*[l]{Computes intersection count. $A$ and $B$ are sorted.}
  $a \leftarrow 0; \quad  b \leftarrow 0; \quad c \leftarrow 0;$ \;
  \While{ $a<|A|$ {\bf and} $b<|B|$}{
    \uIf{$A[b] = B[b]$}{
      ++$c$; \quad ++$a$; \quad ++$b$; \;
    }\ElseIf{$A[a] < B[b]$}{
      ++$a$ \;
    }\Else{
      ++$b$ \;
    }
  }

  \Return $c$

  \caption{\intersect($A, B$)}
  \label{alg:si}
\end{algorithm}

Alg.~\ref{alg:si} outputs the number of the common elements in given two sorted
lists using the set intersection. This algorithm can be used in triangle counting
if the neighborhood of each vertex is sorted based on the vertex IDs. While this
algorithm is cache-friendly, it results in poor performance when there are high
degree vertices. Alg.~\ref{alg:tcsi} outputs the number of triangles in the graph
using this list-based intersection. Alg.~\ref{alg:tch} uses a hash map to count
the number of common neighbors in two vertices' adjacency list.

\begin{algorithm}[t]
  $\tau \leftarrow 0$ \tcp*{Initialize number of triangles to $0$}

  \For{ {\bf each} $u \in V$}{
    \For{{\bf each} $v \in N(G,u)$}{
      $\tau \leftarrow \tau + $ \intersect($N(G,u), N(G,v))$\;
    }
  }

  \Return $\tau$

  \caption{\tci($G$)}
  \label{alg:tcsi}
\end{algorithm}

Variations of Alg.~\ref{alg:tcsi} and Alg.~\ref{alg:tch} are widely used in many
sequential and parallel works~\cite{Shun15-ICDE,Yasar18-HPEC,Latapy08-TCS}.
Latapy~\cite{Latapy08-TCS} proposed a refinement technique inspired from
{\em ayz-listing} that doesn't require $\theta(n^2)$ space. This refinement
technique applies a hash map based algorithm for high-degree vertices and a
list-based algorithm for low-degree vertices. In this paper, we use Latapy's
algorithm as our sequential baseline since it provides the fastest sequential
execution.

\begin{algorithm}[t]

  $\tau \leftarrow 0$ \tcp*[r]{Initialize number of triangles to $0$}

  \For{ {\bf each} $u \in V$}{
    \For{{\bf each} $v \in N(G,u)$}{
      $H[v] \leftarrow u$\;
    }

    \For{{\bf each} $v \in N(G,u)$}{
      \For{{\bf each} $w \in N(G,v)$}{
        \If{ $H[w]=u$}{
          $\tau \leftarrow \tau+1$\;
        }
      }
    }
  }

  \Return $\tau$

  \caption{\tch($G$)}
  \label{alg:tch}
\end{algorithm}

\begin{figure}[ht]
  \begin{minipage}{0.45\columnwidth}
    \subfigure[Toy Graph]{\includegraphics[width=.85\columnwidth]{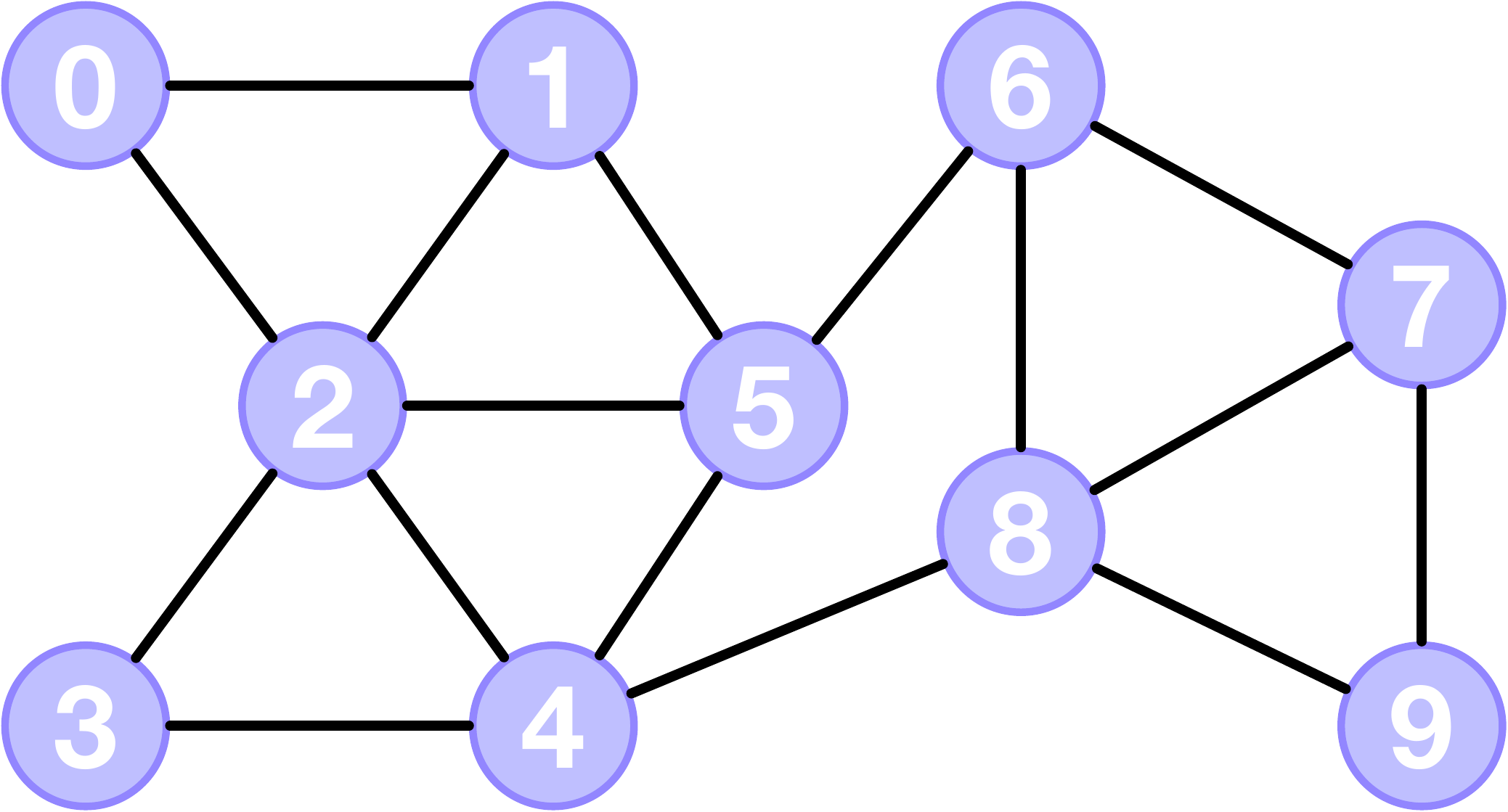}
    \label{fig:toy}}
    \\
    \subfigure[Degree Based Ordering]{\includegraphics[width=.85\columnwidth]{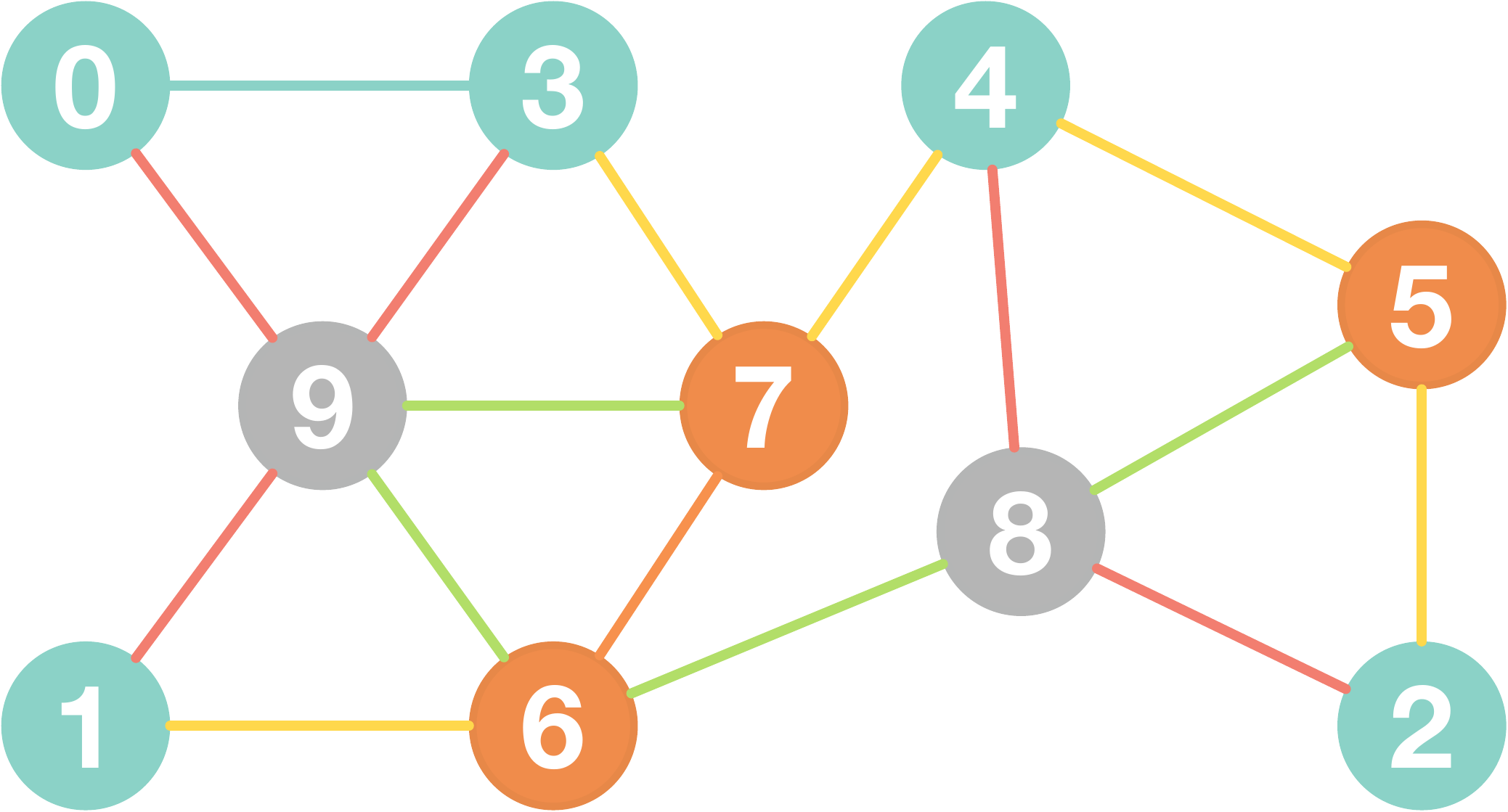}
    \label{fig:toy-deg}}
  \end{minipage}
  \begin{minipage}{0.53\columnwidth}
    \subfigure[Matrix Form]{\includegraphics[width=\columnwidth]{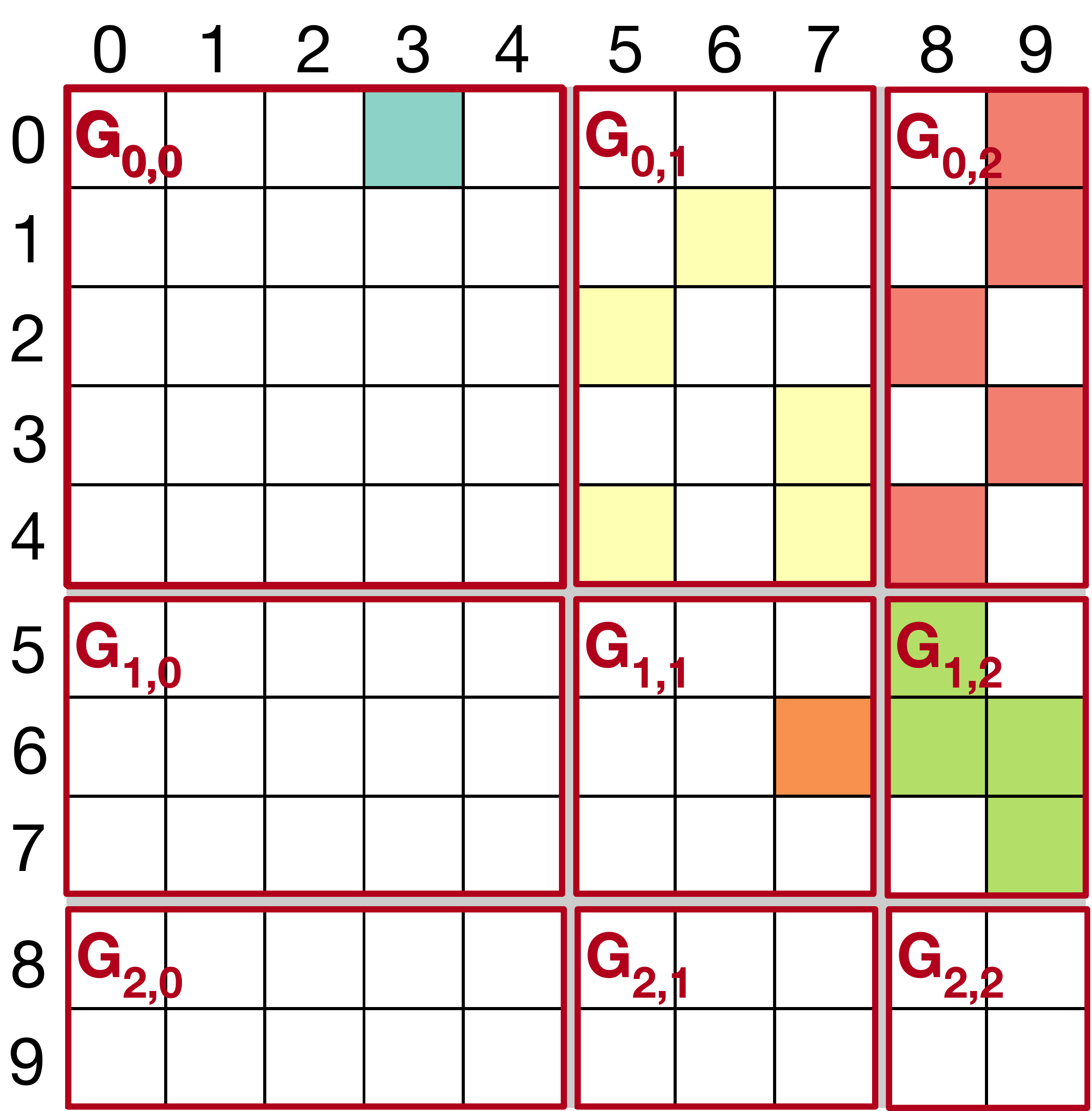}
    \label{fig:toy-parti-mtx}}
  \end{minipage}

  \subfigure[Block CSR Layout]{\includegraphics[width=\columnwidth]{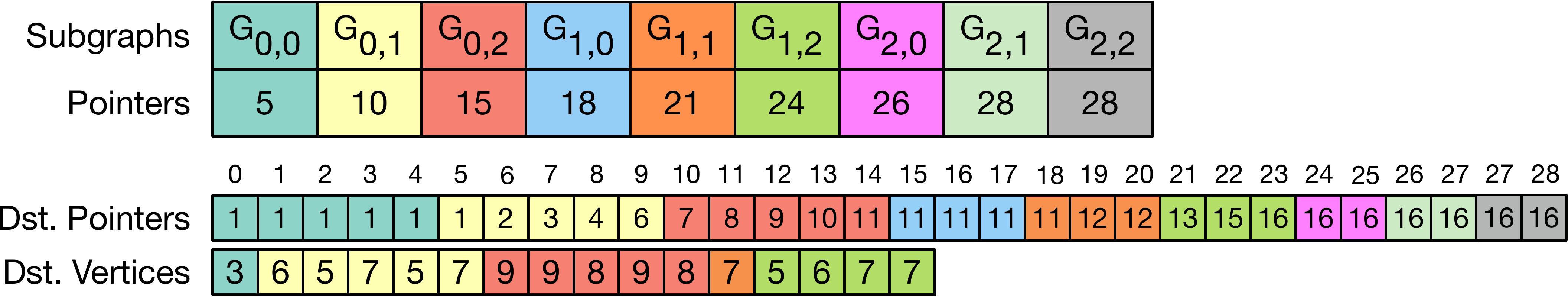}
  \label{fig:toy-layout}}
  \caption{Toy Example. {\bf \ref{fig:toy}}: a toy graph.
    {\bf \ref{fig:toy-deg}}: Degree-based ordering of \ref{fig:toy}, each edge in a block colored with
    same color and each vertex in a vertex partition colored with the same color.
    {\bf \ref{fig:toy-parti-mtx}}: Adjacency matrix representation of the symmetric rectilinear partitioned graph.
    {\bf \ref{fig:toy-layout}}: Block CSR representation of the given partitioned graph.
    Tasks, $T = \{ \{0,0,0\}, \{0,0,1\}, \{0,0,2\}, \{0,1,1\}, \{0,1,2\},\{0,2,2\},$  $\{1,1,1\}
          \{1,1,2\}, \{1,2,2\}, \{2,2,2\} \}$.}
  \label{fig:overview}
\end{figure}

\section{Block-Based Triangle Counting (\bbtc)}
\label{sec:bbtc}

Most of the current state-of-the-art on the triangle counting problem require access
to the entire adjacency list for each vertex. In this paper, we investigate the
usage of partial adjacency lists (i.e., subgraphs).

\subsection{Subgraph generation}

A contiguous rectangle in an adjacency matrix corresponds to an
\emph{edge-induced subgraph} (a subset of edges in the graph).
We propose to use symmetric rectilinear partitioning (also known as symmetric
generalized block distribution~\cite{Grigni96-PAISP}) to \emph{generate} these
subgraphs. This partitioning is
similar to 2D matrix distributions that have been widely used in dense algebra with
cartesian distributions~\cite{Oleary85COMM} (i.e., the same partitioning
vector is used for row and column partition) and also applied to sparse
matrices~\cite{Hendrickson950IJHSC}. In this partitioning, diagonal
blocks are required to be squares.

As stated earlier, in order to avoid double counting and reduce the amount of
edge traversal, we use a total ordering of vertex IDs and also only store
upper triangular part of the adjacency matrix. Furthermore, in order to reduce the
number of edges traversed even further, similar to existing efficient sequential
algorithms~\cite{Latapy08-TCS,Berry14-TCS,Ortmann14-ALENEX,Pagh14-PODS},
we re-order vertices in non-decreasing
degree order, then apply symmetric partitioning to the adjacency matrix after these
transformations. Figure~\ref{fig:overview} illustrates this process on a
toy graph.

Again, by definition, edges of a triangle can appear in at
most three of the subgraphs defined by our block-based partitioning.
Furthermore, in a triangle, for a chosen pair of edges, there exists
only one sub-graph such that the third edge belongs.
Therefore we can bound the computational complexity of triangle
counting and the data movement cost using symmetric rectilinear partitioning.

Our algorithm can work with any valid symmetric partitioning. The only
constraint is the limited memory of the computing devices that will be used.
Therefore, ideal symmetric partitioning should limit the sizes of subgraphs, such
that three subgraphs can fit into memory of the computing devices.
In this work, we used a lightweight symmetric rectilinear partitioning algorithm called
PBD~\cite{Yasar19-arXiv}. After the generation of the blocks, we use a block
variant of CSR (BCSR)~\cite{Im04-JHPCA} to store the graph. This storage format
is illustrated in Fig.~\ref{fig:toy-layout}. In this layout, blocks are ordered
in row-major order.

\subsection{Composition of the task list}

Assume that vertices $u$, $v$ and $w$ form a triangle in a given graph (see Fig.~\ref{fig:tri-example}).
Let the first edge, $(u,v)$, appear in the subgraph $G_{i,j}$. It means that
$u\in V_i$ and $v \in V_j$. Hence, the second edge, $(v,w)$, may appear
in any subgraph $G_{j,k}$ where $j\leq k \leq p-1$ so $w \in V_k$.
Therefore, by definition the third edge, $(u,w)$ appears in $G_{i,k}$ because $u\in V_i$
and $w \in V_k$. In the context of this paper a task can be
defined as below.

\begin{definition}{Task.}
Given a symmetric rectilinear partitioned graph, $G=(V,E)$ that is partitioned
into $p \times p$ blocks, and a triplet $\{ i, j, k\}$ such that $i \leq j \leq k$, a task,
$t$, is defined as the unit of work to count the number of triangles
in $\{G_{i,j}, G_{j,k}, G_{i,k}\}$.
\end{definition}

To illustrate, in Fig.~\ref{fig:toy-deg}, vertices
$3$, $7$ and $9$ form a triangle in the toy graph. The first edge, $(3,7)$,
appears in the subgraph $G_{0,1}$, the second edge, $(7,9)$, appears in
the subgraph $G_{1,2}$ and the third edge, $(3,9)$ appears in the
subgraph $G_{0,2}$.  This would be counted when $G_{0,1}$, $G_{1,2}$ and $G_{0,2}$ form a task;
$t=\{ 0, 1, 2\} = \{G_{0,1}, G_{1,2}, G_{0,2}\}$.

One can count number of triangles, $(u,v,w)$, in given partitioned graph
by considering all possible triples of blocks (i.e., tasks); $\{G_{i,j}, G_{j,k}, G_{i,k}\}$
where $(u,v)\in G_{i,j}$, $(v,w)\in G_{j,k}$ and $(u,w)\in G_{i,k}$.
However in an undirected graph this approach counts each triangle six times.
Most state-of-the-art work treats undirected graphs
as directed by considering edges from smaller vertex IDs to larger
vertex IDs (or vice versa) to be able to count each triangle once. In the block-based
context, this rule can be applied by considering subgraph indices in increasing
order, such that $i \leq j \leq k$.
With this restriction, if the vertex set, $V$, of  a given graph
partitioned into $p$ parts, then number of tasks, $|T|$, is defined
as: $$|T| = \frac{p \times (p+1) \times (p+2)}{6}$$

For instance, as shown in the toy example presented in Fig.~\ref{fig:overview}
$10$, tasks can be defined.
Alg.~\ref{alg:taskcomposition} illustrates task list composition for a given
symmetric rectilinear partitioned graph.

\begin{algorithm}[t]
  $T \leftarrow \emptyset$ \tcp*[r]{Initialize empty task list}

  \For{ $i=0$ {\bf to} $p-1$}{
    \For{ $j=i$ {\bf to} $p-1$}{
      \For{ $k=j$ {\bf to} $p-1$}{
        $T \leftarrow T \cup \{G_{i,j},G_{j,k},G_{i,k}\}$
      }
    }
  }

  \Return $T$

  \caption{\bbtask($G$, $p$)}
  \label{alg:taskcomposition}
\end{algorithm}

\subsection{Counting triangles in a task}
\begin{algorithm}[ht]
  $\tau \leftarrow 0$  \tcp*[r]{Initialize number of triangles to $0$}

  \For{ {\bf each} $u \in V_{S}(G_{i,j})$}{
    \For{{\bf each} $v \in N(G_{i,j},u)$}{
      $\tau \leftarrow \tau+$ \intersect($N(G_{i,k},u), N(G_{j,k},v))$\;
    }
  }

  \Return $\tau$

  \caption{\bbtci($G_{i,j}$, $G_{j,k}$, $G_{i,k}$)}
  \label{alg:bbtcsi}
\end{algorithm}
In the context of this paper a task is defined as counting the number of
triangles that appear in three subgraphs; $G_{i,j}$, $G_{j,k}$ and
$G_{i,k}$ where $i \leq j \leq k$. For each edge, $(u,v)$ in $G_{i,j}$,
this operation can be done by counting the number of common neighbors
between the partial neighbor list of $u$ in $G_{i,k}$ ($N(G_{i,k}, u)$) and the
partial neighbor list of $v$ in $G_{j,k}$ ($N(G_{j,k}, v)$).
Similar to
regular triangle counting algorithms this can be computed using list
or hashmap-based intersection algorithms as
described in Alg.~\ref{alg:bbtcsi} and Alg.~\ref{alg:bbtch} respectively.
\begin{algorithm}[ht]
  $\tau \leftarrow 0$ \tcp*[r]{Initialize number of triangles to $0$}

  \For{ {\bf each} $u \in V_{S}(G_{i,j})$}{
    \For{{\bf each} $v \in N(G_{i,k},u)$}{
      $H[v] \leftarrow u$\;
    }
    \For{{\bf each} $v \in N(G_{i,j},u)$}{
      \For{{\bf each} $w \in N(G_{j,k},v)$}{
        \If{ $H[w]=u$}{
          $\tau \leftarrow \tau+1$\;
        }
      }
    }
  }

  \Return $\tau$

  \caption{\bbtch($G_{i,j}$, $G_{j,k}$, $G_{i,k}$)}
  \label{alg:bbtch}
\end{algorithm}
\subsection{Computational Complexity}

In this subsection, we will present the complexity analysis of \bbtc.
Let the input graph $G=(V,E)$ be partitioned with a $p\times p$ symmetric rectilinear
partitioning. Let $m_{\text{max}}$ and
$m_{\text{avg}}$ be, respectively, the maximum and average number of edges (nonzeros) within subgraphs (blocks).
We also define {\em load imbalance}, $\lambda$, as $\lambda= \frac{m_{\text{max}}}{m_{\text{avg}}}$.
Let $d_{\text{max}}$ be the maximum degree in the input graph, i.e.,
  $$d_{\text{max}} = \max_{u\in V}d(G,u),$$
and $d_{\text{max}}'$ be the maximum degree within all blocks, i.e.,
  $$d_{\text{max}}' = \max_{u \in V_s(G_{i,j}). \forall 0\leq i, j <p} d(G_{i,j}, u).$$
(Please note that, they can be degrees of different vertices).
We define $c$ as the ratio of these max degrees, i.e., $c =\frac{d_{\text{max}}}{d_{\text{max}}'}$.
Clearly $ 1 \leq c \leq p$.

One may see Alg.~\ref{alg:bbtcsi} and Alg.~\ref{alg:bbtch} as iterating over the
edges in $G_{i,j}$ and counting common neighbors of source and destination vertices'
partial neighbor lists that appear in $G_{i,k}$ and $G_{j,k}$, respectively. Hence,
since there can be at most $m_{\text{max}}$ edges in $G_{i,j}$ and maximum degree of a vertex
in a block is defined with $d_{\text{max}}'$, a single task can be solved in $O( m_{\text{max}} d_{\text{max}}' )$
time.
The maximum number of nonzeros ($m_{\text{max}}$) among the blocks can be defined using
load imbalance as, $m_{\text{max}} = \lambda m_{\text{avg}}$.
There are $\frac{p(p+1)}{2}$ blocks, hence the average number of nonzeros ($m_{\text{avg}}$) per block
is $m_{\text{avg}} = \frac{2m}{p(p+1)}$.
When we replace $m_{\text{avg}}$ with that representation, $m_{\text{max}}$ equals to $\lambda \frac{2m}{p(p+1)}$.
A task can be solved in $$O\left(\lambda \frac{2m}{p(p+1)} \frac{d_{\text{max}}}{c}\right)$$ time.
Since there can be at most $$\frac{p (p+1) (p+2)}{6}$$ concurrent tasks, our
block-based triangle counting problem can be solved in
$$O\left( \frac{p (p+1) (p+2)}{6} \lambda \frac{2m}{p(p+1)} \frac{d_{\text{max}}}{c} \right)$$
which is equal to $$O\left( \frac{\lambda p}{c} m d_{\text{max}}\right)$$ when we apply proper
simplifications. When the vertices are ordered based on their degrees,
the intersection operation between neighbor lists of two vertices can be computed in
$O(\sqrt{m})$~\cite{Latapy08-TCS} instead of $O(d_{\text{max}})$. Hence,
block-based triangle counting, using the degree ordering of vertices, can be solved
in $$O\left(\frac{\lambda p}{c} m^{3/2}\right).$$
Note that, for dense graphs, achieving the perfect load-imbalance ($\lambda \approx 1$)
is straightforward, and with high probability, $c$ and $p$ would be nearly equal.
Hence, in dense instances, the computational complexity of the block-based formulation
is the same as the work-optimal triangle counting algorithms.
In irregular problem instances, the load imbalance is crucial in the computational complexity.
Therefore, balanced partitioning of the graph is essential. When the partitioned adjacency matrix
has the perfect load imbalance, the computational complexity of the irregular problem instances
becomes nearly equal to the work-optimal cases.

\section{Parallel and Heterogeneous Execution}
\label{sec:overview}

The \bbtc algorithm can be parallelized by concurrently executing the tasks. The
final number of triangles is the sum of the triangles found on all the
tasks. In this work, we utilize CPUs and GPUs to process all the tasks. At a
high level, there are no architecture-specific changes in the
algorithm. That allows the execution of our algorithm on any accelerator and
CPU combination. However, task handling differs between CPU and GPU because of
architectural differences. While a CPU thread executes a single task, threads
execute cooperatively to complete a task in parallel on GPUs.
Alg.~\ref{alg:bbtc} illustrates this difference.

\begin{figure}[ht]
  \centering
  \includegraphics[width=.95\linewidth]{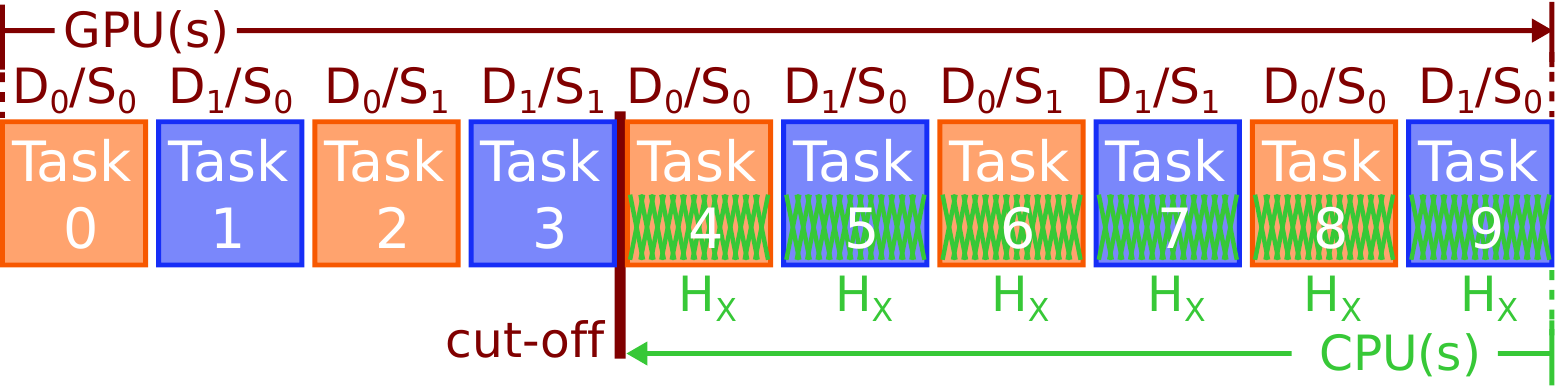}
  \caption{Task Scheduling Between Devices and Host. $D_i$: $i^{th}$
    Device, $S_i$: $i^{th}$ Stream on a Device, $H_x$: a CPU }
  \label{fig:hybrid}
\end{figure}

In hybrid execution environments, assigning computationally heavy tasks to GPUs and lighter
tasks to CPUs (e.g., \cite{Teodoro10-HPDC}) is one of the essential optimizations to
leverage the massive parallelism available on GPUs.
We implement a similar approach.
Tasks are ordered based on their estimations in the task queue ($Q$). GPUs start with heavy tasks
from one end of the execution queue, and CPUs process light tasks from the other end of
the execution queue. They continue to move towards each other as they complete the
assigned tasks. \bbtc keeps a global pointer, ($q_c$), for the last task id that a
CPU thread starts to execute. Each CPU thread atomically decrements the pointer
until finding an unprocessed task or terminates if the counter reaches the
cut-off point of the task queue. That continues while the execution queue is not empty.
Fig.~\ref{fig:hybrid} illustrates this procedure. The execution of tasks
by GPUs and CPUs are described in Alg.~\ref{alg:tgpu} and Alg.~\ref{alg:tcpu},
respectively.

For sorting, we estimate the workload of a task, $t$, using:
$$ExecTime(t) = |\{ (u,v) \in G_{i,j}\}| \times \max\{ \delta(G_{i,k}), \delta(G_{j,k}\}$$
where $\delta(G_{i,j})$
represents the average degree of a subgraph $G_{i,j}$
i.e., $$ \delta(G_{i,j}) = \frac{\sum_{\forall u \in V_i} d(G_{i,j}, u)}{|V_i|}.$$
We sort tasks in the non-increasing order based on their
estimation weights. To decrease the initial synchronization cost of GPUs
and guarantee the assignment of heavier tasks to the GPUs, a cut-off (red
line in Figure~\ref{fig:hybrid}) is pre-defined. CPUs do not go past the
cut-off.

We use CUDA streams to execute several tasks on the GPUs simultaneously.
We create four CUDA streams, ($n_s=4$), for one of the $n_g$ GPUs.
Then, one of the $n_c$ CPU threads is assigned to a stream (see Alg.~\ref{alg:bbtc}).
That CPU thread is responsible for waiting on that stream, synchronizing the stream,
sending a task to a device through that stream, and gathering information from a
device through that stream. All of these operations use
asynchronous function calls. When we create streams and assign a thread
for each of them, GPUs and CPUs compete for tasks and get a new one from the queue
when they finish executing a task.

The cut-off allows all GPU streams to run tasks until the cut-off without
worrying about the next assignment (the first while loop in Alg.~\ref{alg:tgpu}).
GPUs can go past the cut-off line if
more tasks are available (the second while loop in Alg.~\ref{alg:tgpu}).

\begin{algorithm}[ht]
  $gId \leftarrow tId \mod n_s$ \tcp*[r]{GPU id}
  $sId \leftarrow tId / n_s$\tcp*[r]{Stream id}

  \While{$tId<$ cut-off}{
    \For{{\bf each} $G_{i,j}\in T[ tId ]$}{
      \If{{\bf not} {\sc isCopied} ($G_{i,j}, gId$)}{
        {\sc asyncCopy} ($G_{i,j}, gId$)\;
      }
    }
    $G_{i,j}, G_{j,k}, G_{i,k} \leftarrow T[tId]$ \tcp*[r]{Get copied blocks}
    \eIf{{\sc isDense} ($T[tId]$)}{
      {\tt async} \bbtch($G_{i,j}, G_{j,k}, G_{i,k}$)\;
    }{
      {\tt async} \bbtci($G_{i,j}, G_{j,k}, G_{i,k}$)\;
    }
    $tId \leftarrow tId + n_g \times n_s$\;
  }

  \tcp*[l]{Copy blocks for next task, $tId$.}
  \For{{\bf each} $G_{i,j}\in T[ tId ]$}{
    \If{{\bf not} {\sc isCopied} ($G_{i,j}, gId$)}{
      {\sc asyncCopy} ($G_{i,j}, gId$)\;
    }
  }

  \tcp*[l]{Get a task from $Q$ if available, stop otherwise}
  \While{$tId<|T|$ {\bf and not} {\sc atomicSet} ($Q, tId$)}{
      {\sc syncStream} ($sId, gId$) \tcp*[r]{Synchronize the stream}
      $G_{i,j}, G_{j,k}, G_{i,k} \leftarrow T[tId]$ \tcp*[r]{Get copied blocks}
      \eIf{{\sc isDense} ($T[tId]$)}{
        {\tt async} \bbtch($G_{i,j}, G_{j,k}, G_{i,k}$)\;
      }{
        {\tt async} \bbtci($G_{i,j}, G_{j,k}, G_{i,k}$)\;
      }
    $tId \leftarrow tId + n_g \times n_s$\;
    \For{{\bf each} $G_{i,j}\in T[ tId ]$}{
      \If{{\bf not} {\sc isCopied} ($G_{i,j}, gId$)}{
        {\sc asyncCopy} ($G_{i,j}, gId$)\;
      }
    }
  }

  \caption{\tgpu($tId$)}
  \label{alg:tgpu}
\end{algorithm}

\begin{algorithm}[ht]
  $tId \leftarrow$ {\sc atomicDecrement} ($q_c$) \tcp*[r]{Get task index.}
  $\tau \leftarrow 0$\;
  \While{$t \geq $ cut-off}{
    \tcp*[l]{Check if task has been processed and set if not.}
    \eIf{{\bf not} {\sc atomicSet} ($Q, tId$)}{
      $G_{i,j}, G_{j,k}, G_{i,k} \leftarrow T[tId]$\;
      \eIf{{\sc isDense} ($T[tId]$)}{
        $\tau \leftarrow \tau +$ \bbtch($G_{i,j}, G_{j,k}, G_{i,k}$)\;
      }{
        $\tau \leftarrow \tau +$ \bbtci($G_{i,j}, G_{j,k}, G_{i,k}$)\;
      }
    }{
      $tId \leftarrow$ {\sc atomicDecrement} ($q_c$)\;
    }
  }
  \Return $\tau$

  \caption{\tcpu()}
  \label{alg:tcpu}
\end{algorithm}

We allow GPUs and CPUs to compete for the tasks. However, GPUs do not compete
among themselves. We assign tasks to GPUs in a deterministic round-robin fashion.
This approach allows stream threads to know the next task that can be executed by
that stream. Then a stream thread can overlap copying the blocks of the next
assignment with the computation.

\begin{algorithm}[ht]
  \tcp*[l]{Initialization of the system variables.}
  $n_c \leftarrow $ {\sc nCores}(); \hspace{1ex}
  $n_g \leftarrow $ {\sc nGPUs}(); \hspace{1ex}
  $n_s \leftarrow $ 4;\;
  $\tau \leftarrow 0$ \tcp*[r]{Number of triangles}

  \vspace{.5em}
  \tcp*[l]{Partitioning}
  $G(V^p, E) \leftarrow$ {\sc PBD}($G, p$) \tcp*[r]{Parallel PBD algorithm}

  \vspace{.5em}
  \tcp*[l]{Composing task list}
  $T \leftarrow $ \bbtask($G, p$) \tcp*[r]{Composing task list}
  $T \leftarrow $ {\sc orderTasks} ($T$) \tcp*[r]{Heuristic based ordering of tasks}

  \vspace{.5em}
  \tcp*[l]{Initializing a list of bit flags for each task}
  $Q[i] = 0$, for $0 \leq i \leq |T|-1$ \;

  $q_c \leftarrow |T|$ \tcp*[r]{Queue index for competing cores.}

  \vspace{.5em}
  \tcp*[l]{Spawning gpu threads: one per each stream}
  \For{$i=0$ {\bf to } $n_s-1$}{
    \For{$j=0$ {\bf to} $n_g-1$}{
      {\tt spawn} \tgpu($i\times s + j$)\;
    }
  }

  \vspace{.5em}
  \tcp*[l]{Spawning cpu threads}
  \For{$i=n_s\times n_g$ {\bf to} $n_c$}{
    {\tt spawn} \tcpu()\;
  }

  \vspace{.5em}
  \tcp*[l]{Gathering \& aggregating counts}
  \For{$i=0$ {\bf to } $n_s-1$}{
    \For{$j=0$ {\bf to} $n_g-1$}{
      $t \leftarrow$ {\sc getCount}($i,j$)\;
      {\sc atomicAdd} ($\tau, t$)\;
    }
  }

  \Return $\tau$

  \caption{\bbtc($G, p$)}
  \label{alg:bbtc}
\end{algorithm}

\section{Related Work}
\label{sec:related}

Inspired by {\em AYZ listing}~\cite{Alon97-Algorithmica}, Latapy~\cite{Latapy08-TCS} proposed a refinement
technique that does not require $\theta(n^2)$ space. Latapy's algorithm uses
the list-based intersection for small degree vertices and the hashmap-based
intersection for high degree vertices. In this work, we use a similar approach; i.e.,
any task with sparse blocks uses the list-based intersection, and any task consist of dense
blocks uses a hashmap. Shun and Tangwongsan~\cite{Shun15-ICDE} parallelized Latapy's
{\em compact-forward} algorithm. Recently, Dhulipala et al.~\cite{Dhulipala18-SPAA}
re-implemented an optimized version of this algorithm.

A linear-algebra-based triangle counting implementation, kkTri (previously
designated TCKK), was proposed by Wolf et al.~\cite{Wolf17-HPEC}. That work
focused on efficient shared memory parallelism on top of a portable SpGEMM
(called KK-MEM)~\cite{Deveci17-IPDPSW} in the Kokkos Kernels
library\footnote{\url{https://github.com/kokkos/kokkos-kernels}}. This work
optimized two linear-algebra-based formulations of the triangle counting problem
and used different types of hashmap accumulators based on the sparsity of the graphs.
In a later work, Ya\c{s}ar et al.~\cite{Yasar18-HPEC} improved this algorithm by using
a 1D task-based approach and an adaptive runtime. In this paper, we use 2D
partitioning instead of 1D partitioning, and we always use dense hashmaps.
The memory requirement for dense hashmaps decreases significantly because of
the 2D partitioning. The smaller dense hashmaps are also more suitable for GPU architectures.

Green et al.~\cite{Green14-WIAP} proposed a list-intersection-based algorithm on
a single GPU. In that work, neighbor lists of source and destination vertices of
each edge are concurrently processed by $32$ threads from a warp. However, this
approach comes with expensive partitioning overhead.
Hu et al.~\cite{Hu18-SC} try to overcome this problem by introducing a binary
search-based intersection method. Both works copy the graph into GPU
memory before starting the computation, which causes significant
efficiency issues because of the idle time. Besides, using that approach~\cite{Green14-WIAP},
one cannot process graphs that are larger than the GPU's memory size.
\cite{Hu18-SC} addresses this problem by applying a rectilinear partitioning
strategy; however, their execution time is drastically affected by the number of
partitions. Date et al.~\cite{Date17-HPEC} proposed to use CPUs and GPUs in a
collaborative fashion using GPU zero-copy memory, aiming to decrease CPU-GPU
data transfer overhead by leveraging unified memory capabilities. Inefficient
use of GPU memory causes drastic performance decrease in that case.

Recently, Tom et al.~\cite{Tom19-ICPP} and Acer et al.~\cite{Acer19-HPEC} proposed
distributed-memory triangle counting algorithms. Tom et
al.~\cite{Tom19-ICPP} propose a 2D distributed-memory triangle counting algorithm
that follows similar steps to Cannon's parallel matrix multiplication algorithm.
Acer et al.~\cite{Acer19-HPEC} propose an MPI based distributed memory parallel
2D triangle counting algorithm which tries to exploit the benefits of shared
memory parallelism using Cilk.

\section{Experimental Evaluation}
\label{sec:exp}

We present several experiments to identify the performance trade-offs of the
proposed work. In our tests, we used three architectures with
multi-core processors and multi-GPUs. Table~\ref{table:hardware} lists the properties of those
architectures.
Depending on the architecture GNU compiler (g++) version $7.2$ or Intel compiler
version $19.03$ (icpc), CUDA runtime version $10.0$ and OpenMP version
$4.0$ are used to compile and run the code. The source code of BSD-licensed \bbtc is available at
\url{http://tda.gatech.edu/software/bbtc/}.

\setlength{\tabcolsep}{1.7pt}
\begin{table}[t]
\caption{Overview of the Architectures. Pinned: transfers through pinned
memory. Pageable: transfers through pageable memory. Peer to peer: transfers
done between devices. H2D: host to device. D2H: device to host. (measured bandwidth)}
\label{table:hardware}
\begin{center}
\begin{tabular}{| l l | l | l | l |}
\hline
  &  & \textbf{DGX} & \textbf{Newell} & \textbf{Haswell}\\
\hline

\multicolumn{2}{|l|}{\textbf{CPU}}
  & Intel, E5-2698
  & POWER9
  & Intel, E7-4850\\ 

\multicolumn{2}{|l|}{\textbf{Cores}}
  & $2 \times 20$
  & $2 \times 16$
  & $4 \times 14$\\ 

\multicolumn{2}{|l|}{\textbf{Host Memory}}
  & 512 GB
  & 320 GB
  & 2 TB\\ 

\multicolumn{2}{|l|}{\textbf{L2-Cache}}
  & 256 KB
  & 512 KB
  & 256 KB \\ 

\multicolumn{2}{|l|}{\textbf{L3-Cache}}
  & 50 MB
  & 10 MB
  & 35 MB\\ \hline \hline

\multicolumn{2}{|l|}{\textbf{GPU}}
  & V100
  & V100
  & N/A\\ 

\multicolumn{2}{|l|}{\textbf{GPU Memory}}
  & 32 GB
  & 32 GB
  & N/A\\ 

\multicolumn{2}{|l|}{\textbf{Number of GPUs}}
  & 8
  & 2
  & N/A\\ \hline \hline

\multirow{2}{*}{\textbf{Pageable}}
    & H2D     & 9.2 GB/s &  12.2 GB/s &  N/A\\
    & D2H     & 8.0 GB/s &  14.2 GB/s &  N/A\\ 

\multirow{2}{*}{\textbf{Pinned}}
    & H2D     & 10.7 GB/s & 60.0 GB/s &  N/A\\
    & D2H     & 12.1 GB/s & 60.0 GB/s &  N/A\\ 

\multicolumn{2}{|l|}{\textbf{Peer to Peer}}
  & 23.5 GB/s
  & 31.3 GB/s
  & N/A\\ \hline

\end{tabular}
\end{center}
\end{table}

\subsection{Used state-of-the-art works for comparison}
\label{ssec:stateofart}

\renewcommand*{\thefootnote}{\fnsymbol{footnote}}
\setlength{\tabcolsep}{1.7pt}
\begin{table}[t]
\caption{Overview of Design Choices. $^\dag$ Complexity for one dimensional parallel case. Complexity of partitioned version is not provided in~\cite{Hu18-SC}.}
\label{table:overview}
\begin{center}
\begin{tabular}{lcccc}
\hline
 & \multicolumn{3}{c}{State of The Art} & \\
\cline{2-4}
                           & TCM   & kkTri   & TriCore & \bbtc\\
\hline \hline
\multicolumn{5}{c}{Algorithmic properties} \\ \hline \hline
       List Intersect.     & \cmark & \xmark & \xmark & \cmark \\
       Map Intersect.  & \xmark & \cmark & \xmark & \cmark \\
       Search Intersect.     & \xmark & \xmark & \cmark & \xmark\\
       Multi-Core            & \cmark & \cmark & \xmark & \cmark\\
       Multi-GPU             & \xmark & \xmark & \cmark & \cmark\\
       Complexity     & $O(m^{\frac{3}{2}})$
                       & $O(m^{\frac{3}{2}})$
                       & $O(m^{\frac{3}{2}} \log{\sqrt{m}})$$^\dag$
                       & $O(\frac{\lambda p}{c} m^{\frac{3}{2}})$ \\

\hline \hline
\multicolumn{5}{c}{Implemented optimizations} \\ \hline \hline
       Vertex Ordering       & \cmark & \cmark & \cmark & \cmark\\
       Compression           & \cmark & \cmark & \xmark & \xmark\\
       CUDAStreams           & N/A & N/A & \xmark & \cmark\\
\hline \hline
\multicolumn{5}{c}{Parallelization strategy} \\ \hline \hline
       One-dimension        & \cmark & \cmark & \cmark & \xmark\\
       Two-dimension  & \xmark & \xmark & \cmark & \cmark\\

\hline \hline
\multicolumn{5}{c}{Used runtimes for implementation} \\ \hline \hline
       Pthread based              & \cmark & \xmark & \xmark & \xmark\\
       OpenMP                & \cmark & \cmark & \xmark & \cmark\\
       Cilk                  & \cmark & \cmark & \xmark & \cmark\\
       TBB                   & \xmark & \xmark & \xmark & \cmark\\
       CUDA                  & \xmark & \xmark & \cmark & \cmark\\

\hline
\end{tabular}
\end{center}
\end{table}

\renewcommand*{\thefootnote}{\arabic{footnote}}

In our experiments, we compared \bbtc's performance with three state-of-the-art
implementations: (1) TCM~\footnote{TCM: \url{https://github.com/Ldhulipala/gbbs}},
a multi-core triangle counting algorithm proposed by Shun and
Tangwongsan~\cite{Shun15-ICDE} and optimized by Dhulipala et
al.~\cite{Dhulipala18-SPAA}. We use Dhulipala et al.~\cite{Dhulipala18-SPAA}'s
optimized version.
(2) kkTri~\footnote{kkTri: \url{https://github.com/Kokkos/kokkos-kernels}}, a
multi-core linear algebra-based triangle counting algorithm proposed by Wolf et
al.~\cite{Wolf17-HPEC} and improved by Ya\c{s}ar et al.~\cite{Yasar18-HPEC}.
We use the optimized version on Intel architectures.
(3) TriCore~\footnote{TriCore: \url{https://github.com/huyang1988/TC}}, a
multi-GPU triangle counting algorithm implemented by Hu et al.~\cite{Hu18-SC}
and improved in~\cite{Hu18-HPEC}. We use TriCore's optimized (latest) version~\cite{Hu18-HPEC}.
While TCM and kkTri algorithms have the optimal complexity, TriCore and \bbtc
algorithms have higher complexities due to their design choices. Note that,
the complexity of TriCore is for the one-dimensional parallel case. TriCore also implements a
variant of rectilinear partitioning~\cite{Grigni96-PAISP} to be able to process
graphs that cannot fit into GPU memory. This partitioning increases the cost of
their algorithm. However, the complexity of TriCore's partitioned version is not
provided in~\cite{Hu18-SC}. Table~\ref{table:overview} presents a high-level
overview of the design choices of these algorithms.

\subsection{Dataset and peak rates}

We evaluated the \bbtc algorithm on $16$ real-world, and $8$
synthetic graphs (RMAT) from the SuiteSparse Matrix Collection~\cite{Davis11-TOMS},
SNAP~\footnote{SNAP Datasets: \url{http://snap.stanford.edu/data}},
web data commons~\footnote{WDC Dataset: \url{http://webdatacommons.org/hyperlinkgraph/2014-04}}, and
GraphChallange~\footnote{Graph Challenge Datasets: \url{https://graphchallenge.mit.edu/data-sets}}.
Table~\ref{table:dataset} presents, the properties of these graphs:
The number of vertices ($|V|$), the number of edges
($|E|$), the number of triangles, the compression ratio of the graph (${Comp.}$),
the size of the graph in
memory (Raw Size), the number of tiles, and the number of tasks. For all experiments, we get the
median of five runs with different numbers of GPUs and report the best. For a graph, the rate
is defined as the ratio between the execution time and the number of edges. Best rates
are achieved on the DGX machine. Table~\ref{table:dataset} also reports rates for
each graph.

\setlength{\tabcolsep}{1pt}
\begin{table*}[th]
    \caption{ Properties of the dataset.
    Best of the medians of
    execution times in seconds and corresponding rates are reported.}
    \label{table:dataset}
    \begin{center}
      \begin{tabular}{ | l|| 
          r r r||
          r ||
          r ||
          r | r ||
          r  c |
        }
        \hline

        \multirow{2}{*}{\textbf{Data Set}}  &
        \multirow{2}{*}{$\boldsymbol{|V|}$}   &
        \multirow{2}{*}{$\boldsymbol{|E|}$}   &
        \multirow{2}{*}{\textbf{Triangle Count}}   &
        \multirow{2}{*}{\textbf{Comp.}}  &
        \multirow{2}{*}{\textbf{Raw Size}}  &
        \multirow{2}{*}{\textbf{Tiles}}  &
        \multirow{2}{*}{\textbf{Tasks}}  &
        \multicolumn{2}{c|}{\textbf{Best (Copy Included)}}  \\

        & & & & & & & & \textbf{Time (s)} & \textbf{Rate ($\times 10^8$)}\\ \hline \hline

    cit-HepTh  &  27,770  &  352,285  &  1,478,735
        &  0.6  
        & 2.2 MB
        & 64 & 120
        & 0.002 & 1.7 \\ \hline

    email-EuAll  &  265,214  &  364,481  &  267,313
        &  0.5  
        &  9.5 MB
        & 64 & 120
        & 0.002 & 1.9 \\ \hline

    soc-Epinions1  &  75,879  &  405,740  &  1,624,481
        &  0.6  
        & 3.9 MB
        & 64 & 120
        & 0.002 & 2.1 \\ \hline

    cit-HepPh  &  34,546  &  420,877  &  1,276,868
        &  0.5  
        & 2.7 MB
        & 64 & 120
        & 0.002 & 2.0 \\ \hline

    soc-Slashdot0811  &  77,360  &  469,180  &  551,724
        &  0.6  
        & 5.4 MB
        & 144 & 364
        & 0.002 & 2.9 \\ \hline

    soc-Slashdot0902  &  82,168  &  504,230  &  602,592
        &  0.6  
        & 5.7 MB
        & 144 & 364
        & 0.002 & 3.1 \\ \hline

    flickrEdges  &  105,938  &  2,316,948  &  107,987,357
        &  0.6  
        & 14 MB
        & 144 & 364
        & 0.016 & 1.5 \\ \hline

    amazon0312  &  400,727  &  2,349,869  &  3,686,467
        &  0.7  
        & 28 MB
        & 144 & 364
        & 0.006 & 3.9 \\ \hline

    amazon0505  &  410,236  &  2,439,437  &  3,951,063
        &  0.7  
        & 29 MB
        & 144 & 364
        & 0.007 & 3.7 \\ \hline

    amazon0601  &  403,394  &  2,443,408  &  3,986,507
        &  0.7  
        & 28 MB
        & 144 & 364
        & 0.006 & 4.4 \\ \hline

    scale18  &  174,147  &  3,800,348  &  82,287,285
        &  0.6  
        & 26 MB
        & 256 & 816
        & 0.021 & 1.8 \\ \hline

    scale19  &  335,318  &  7,729,675  &  186,288,972
        &  0.6 
        & 56 MB
        & 400 & 1540
        & 0.041 & 1.9 \\ \hline

    as-Skitter  &  1,696,415  &  11,095,298  &  28,769,868
        &  0.8  
        & 146 MB
        & 400 & 1540
        & 0.027 & 4.1 \\ \hline

    scale20  &  645,820  &  15,680,861  &  419,349,784
        &  0.6  
        & 110 MB
        & 400 & 1540
        & 0.079 & 2.0 \\ \hline

    cit-Patents  &  3,774,768  &  16,518,947  &  7,515,023
        &  0.5  
        & 352 MB
        & 400 & 1540
        & 0.038 & 4.4 \\ \hline

    scale21  &  1,243,072  &  31,731,650  &  935,100,883
        &  0.6  
        & 216 MB
        & 400 & 1540
        & 0.144 & 2.2 \\ \hline

    soc-LiveJournal1  &  4,847,571  &  42,851,237  &  285,730,264
        & 0.7  
        & 534 MB
        & 400 & 1540
        & 0.121 & 3.6 \\ \hline

    scale22  &  2,393,285  &  64,097,004  &  2,067,392,370
        & 0.6  
        & 464 MB
        & 576 & 2600
        & 0.325 & 2.0 \\ \hline

    scale23  &  4,606,314  &  129,250,705  &  4,549,133,002
        & 0.5  
        & 915 MB
        & 576 & 2600
        & 0.549 & 2.4 \\ \hline

    scale24  &  8,860,450  &  260,261,843  &  9,936,161,560
        & 0.5  
        & 1.9 GB
        & 784 & 4060
        & 1.154 & 2.3 \\ \hline

    scale25  &  17,043,780  &  523,467,448  &  21,575,375,802
        & 0.5  
        & 4.0 GB
        & 1024  & 5984
        & 2.400 & 2.2 \\ \hline

    twitter  &  61,578,414  &  1,202,513,046  &  34,824,916,864
        & 0.5  
        & 13 GB
        & 1296  & 8436
        & 4.582 & 2.6 \\ \hline

    friendster  &  65,608,366  &  1,806,067,135  &  4,173,724,142
        & 0.5  
        & 16 GB
        & 1296  & 8436
        & 3.133 & 5.8 \\ \hline

    wdc-2014  &  1,724,573,718  &  124,141,874,032  &  4,587,563,913,535
        & 0.8  
        & 650 GB
        & 1296  & 8436
        & 116.5 & 12.4 \\ \hline

      \end{tabular}
    \end{center}
\end{table*}

\subsection{Sequential execution time for varying partitions}
\label{ssec:lambda}

\begin{figure}[ht]
  \centering
  \subfigure[Friendster]{\includegraphics[width=.48\linewidth]{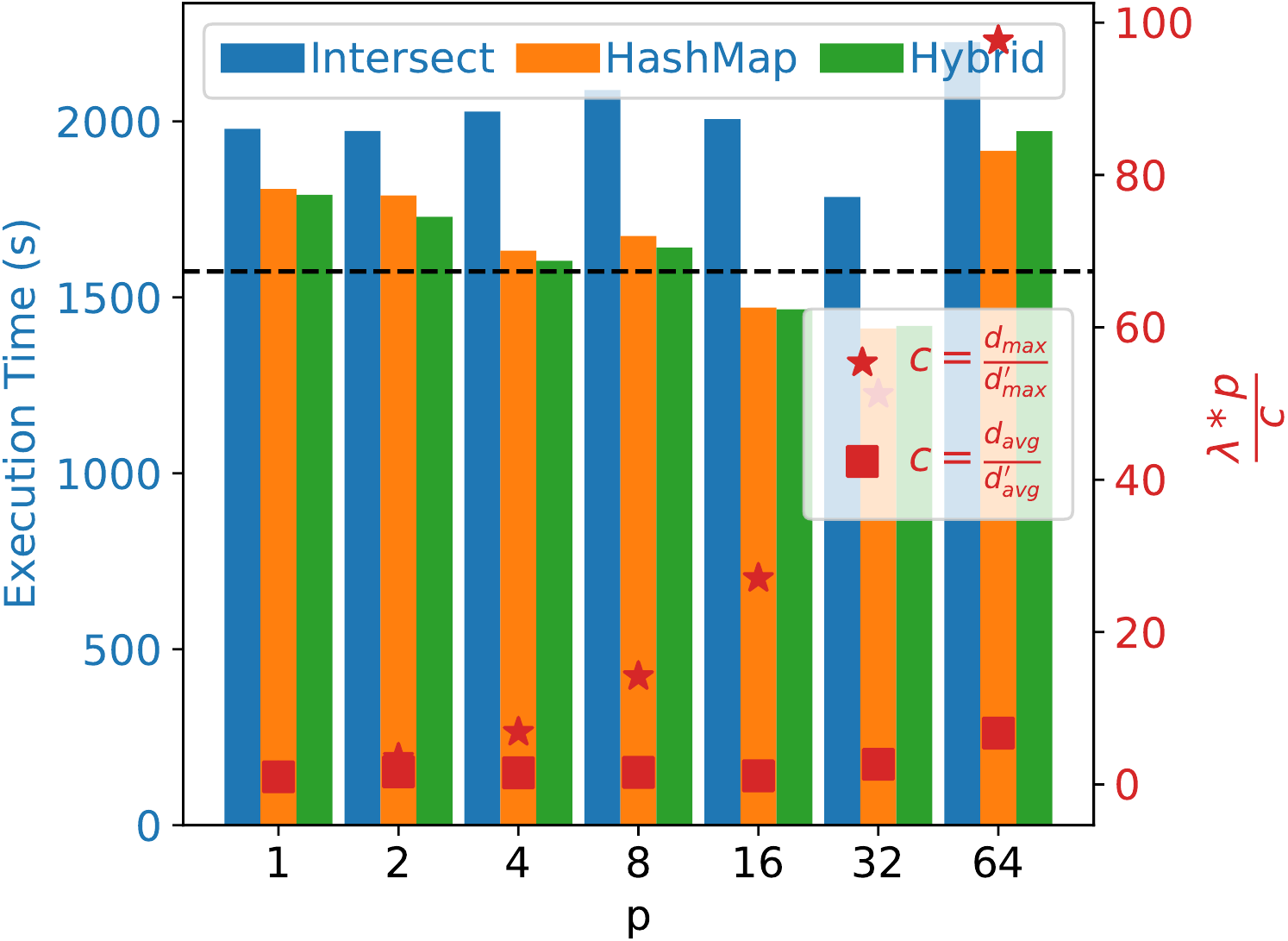}
  \label{fig:friendster-diff-k}}
  \subfigure[Twitter]{\includegraphics[width=.48\linewidth]{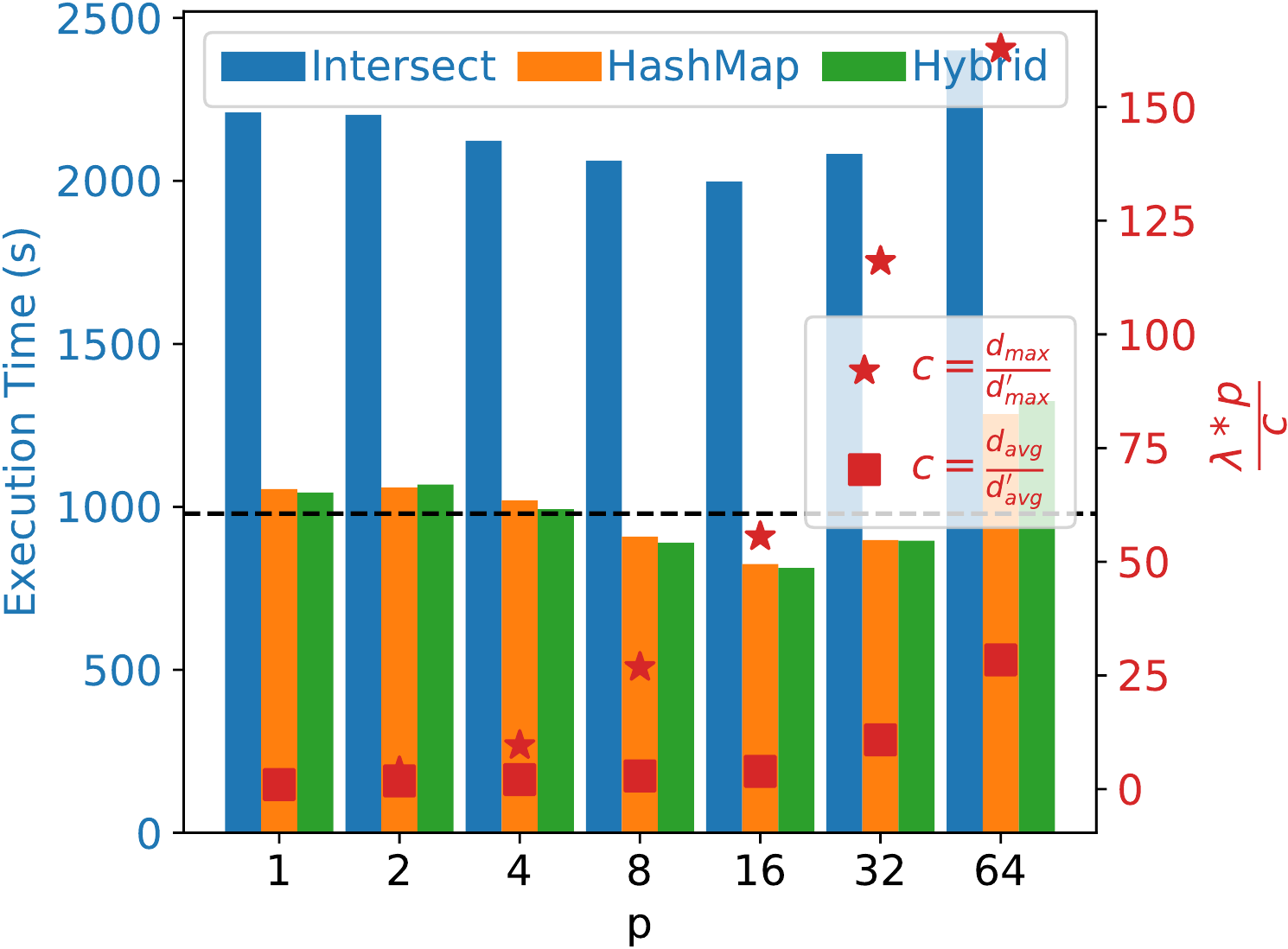}
  \label{fig:twitter-diff-k}}
  \caption{Sequential time of Friendster and Twitter graphs for different $p$ values.
    Black dashed line represents Latapy's algorithm's execution time.}
  \label{fig:asymp}
\end{figure}

Figure~\ref{fig:asymp} shows the sequential behavior of the \bbtc algorithm with respect to
the different number of partitions ($p \times p$) on the Friendster and the Twitter graphs.
Each bar represents the execution time (left-axis) of the \bbtc algorithm
using the list-based intersection, hashmap-based intersection and
hybrid intersection, for a given $p$ (x-axis). The black dashed line represents
the best sequential time that we get using Latapy's algorithm~\cite{Latapy08-TCS}.
We observe that in both graphs  when $p$ is increased,
sequential execution time first decreases
due to better memory utilization and then increases because of
higher load imbalance. Note that, the \bbtc
algorithm runs in $O(\frac{\lambda p}{c}m^{\frac{3}{2}})$ time.
To analyze the behaviour of the \bbtc algorithm, we have plotted
$\frac{\lambda p}{c}$ (right-axis) values for $c=\frac{d_{\text{max}}}{d_{\text{max}}'}$  (star-shaped marker) and
$c=\frac{d_{\text{avg}}}{d_{\text{avg}}'}$  (square-shaped marker) where $d_{\text{max}}$, $d_{\text{avg}}$ are the maximum
and the average degrees in the given graph, and $d_{\text{max}}'$, $d_{\text{avg}}'$ are
the maximum and the average degrees in the blocks of the partitioned graph, respectively.
We observe that trend in execution times are more co-related with
$c=\frac{d_{\text{avg}}}{d_{\text{avg}}'}$.
Hence, one should pick $p$ around the average degree of the graph.
Note that, the \bbtc algorithm outperforms Latapy's algorithm on both graphs
for $p$ values closer to $d_{\text{avg}}$.

\subsection{Comparison with Latapy's sequential algorithm}
\label{ssec:latapycomp}

\begin{figure}[ht]
  \centering
  \subfigure[Perf. Profile of Latapy's Alg.]{\includegraphics[width=.45\linewidth]{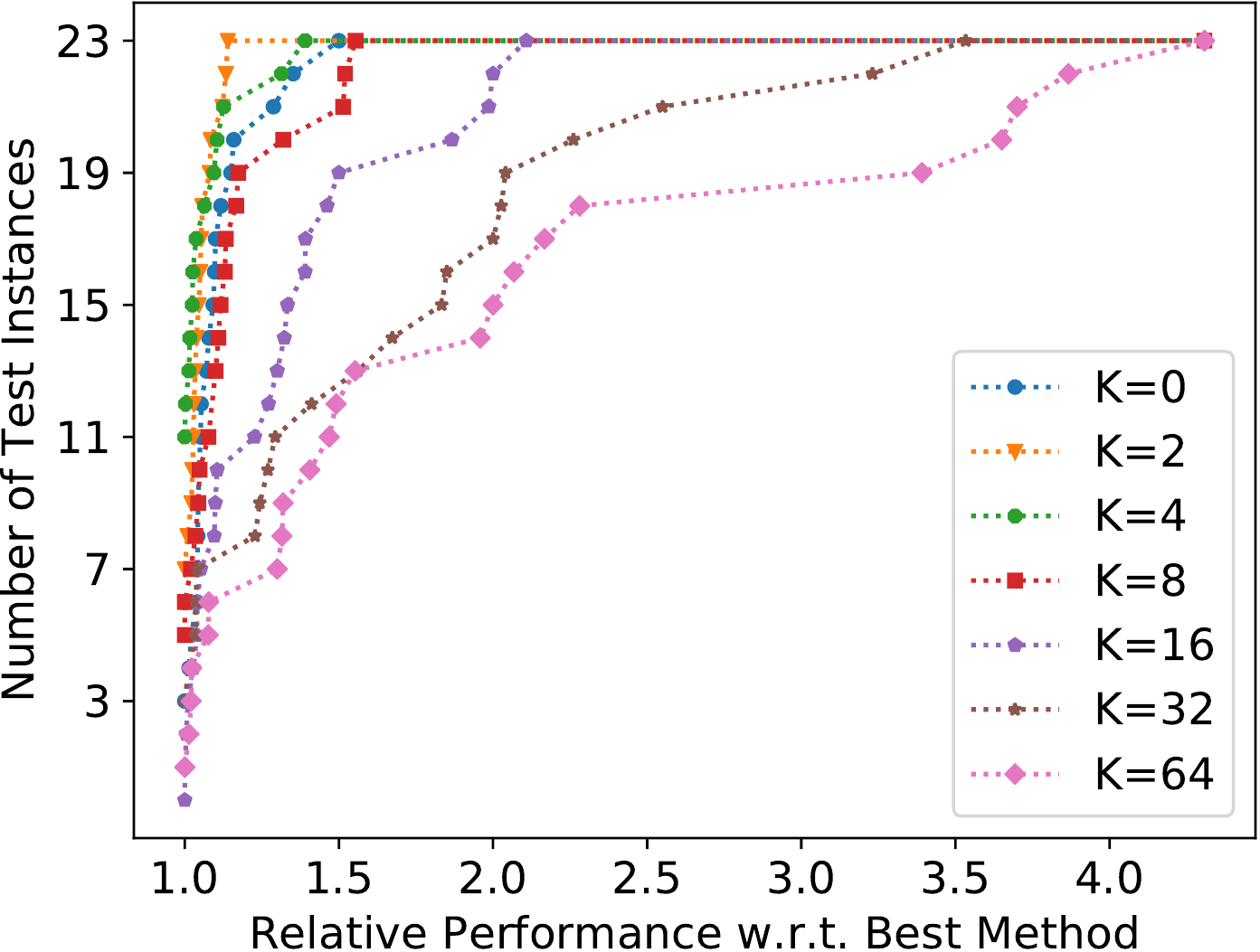}
  \label{fig:latapy-pp}}
  \subfigure[Sequential Execution Times]{\includegraphics[width=.45\linewidth]{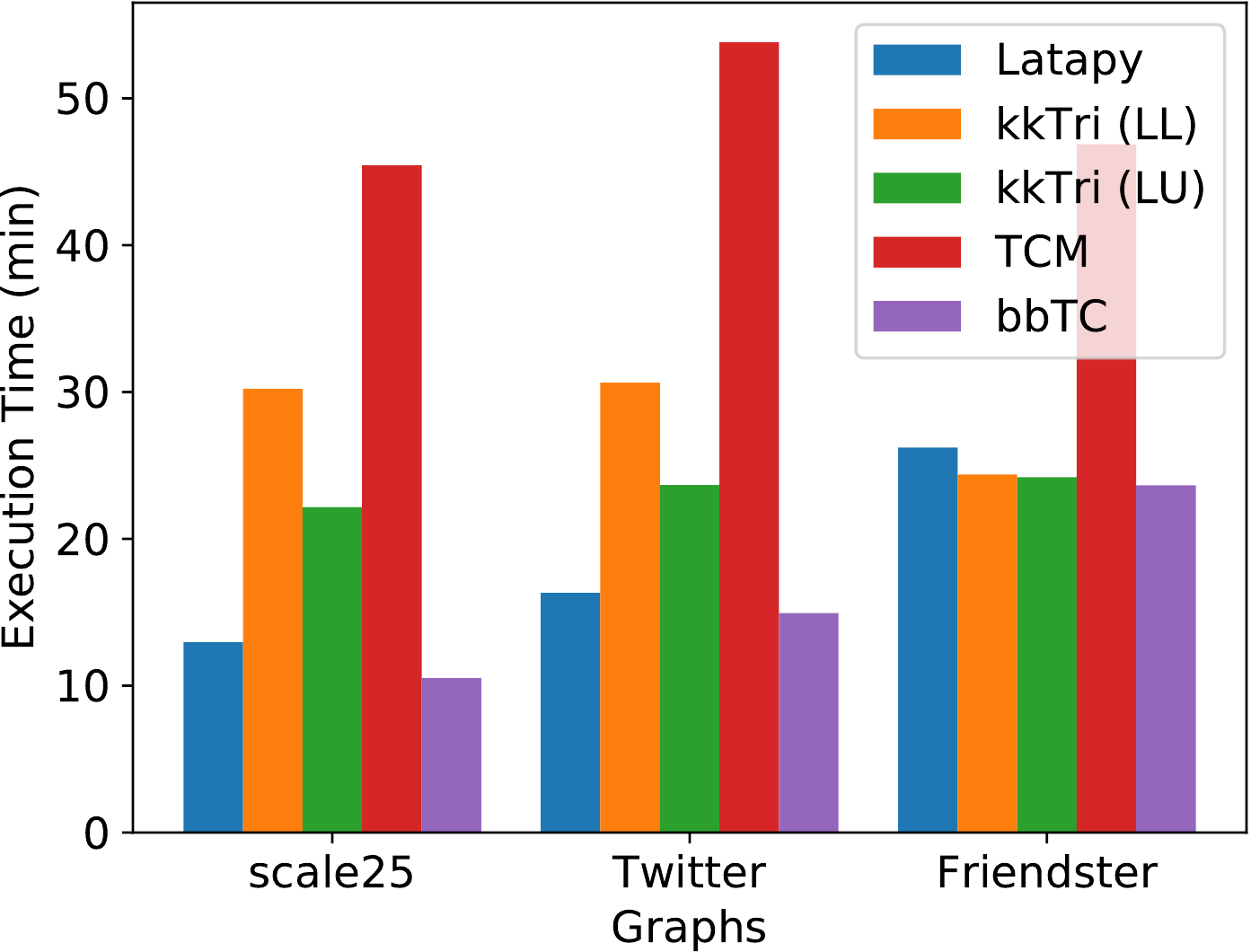}
  \label{fig:seqperf}}
  \caption{Evaluation of Latapy's algorithm wrt. $K$ (\ref{fig:latapy-pp}) and
           comparison of sequential execution times (\ref{fig:seqperf}).}
  \label{fig:seqcomp}
\end{figure}

Given $K$, Latapy's algorithm~\cite{Latapy08-TCS} applies hashmap-based
intersection for vertices whose degrees are higher than $K$, and list-based
intersection for the others. We evaluate Latapy's algorithm's~\cite{Latapy08-TCS}
performance with respect to different $K$ values. Fig.~\ref{fig:latapy-pp} illustrates the
performance profile of the algorithm. In the performance profile, we plot the
number of the test instances (y-axis) in which a $K$ value obtains an execution
time on an instance that is no larger than $x$ times (x-axis) the best execution
time achieved by any $K$ value for that instance~\cite{Dolan02-MP}. Therefore,
the higher a profile at a given $x$ value, the better a $K$ value is. We observe
that switching between list-based and hashmap-based intersection techniques
improves efficiency. In the majority of the test instances, $K=2$ performs better
than others due to the number of smaller graphs in our dataset. In larger
instances, such as Twitter and Friendster, $K=32$ and $K=64$ gives the best
performance. Hence, $K$ should be picked based on the graph size.

Fig.~\ref{fig:seqperf} illustrates sequential execution times of Latapy's
algorithm, kkTri with LL, and LU formulations~\cite{Wolf17-HPEC}, TCM~\cite{Dhulipala18-SPAA},
and \bbtc on the three large graphs of our dataset; scale25, Twitter and Friendster.
\bbtc  outperforms other algorithms in all graph instances due to better memory
utilization. TCM performs the worst because of the list-based intersection.
TCM achieves better scalability in parallel settings with list-based
intersection, because using dense hashmaps (i.e. an array of the length $n=|V|$)
causes poor memory utilization. However, in a sequential setting list-based intersection
performs worse than hashmap-based intersection because there is no bandwidth competition
among running threads.
kkTri uses a linked-list-based hashmap for large graphs, which helps it to outperform Latapy's
algorithm on the Friendster graph. We observe that kkTri's LU formulation
outperforms LL formulation in the sequential case while LL performs better in parallel
setting~\cite{Yasar18-HPEC}.

\subsection{Task Workload Estimation}
\label{ssec:workload}

\begin{figure}[ht]
  \centering
  \subfigure[Friendster; heaviest tasks.]{\includegraphics[width=.45\linewidth]{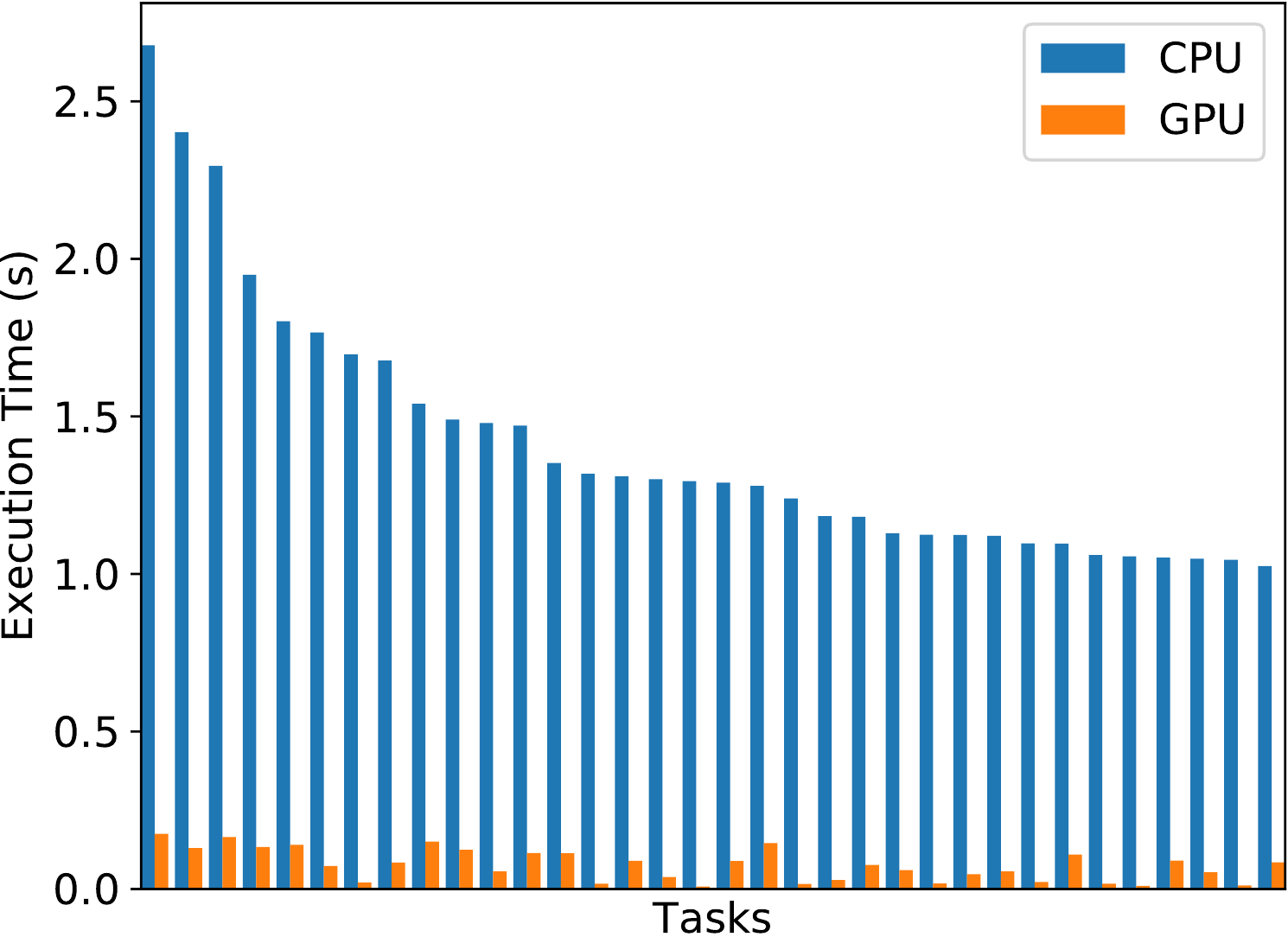}
  \label{fig:taskrun-fri}}
  \subfigure[Twitter; heaviest tasks.]{\includegraphics[width=.45\linewidth]{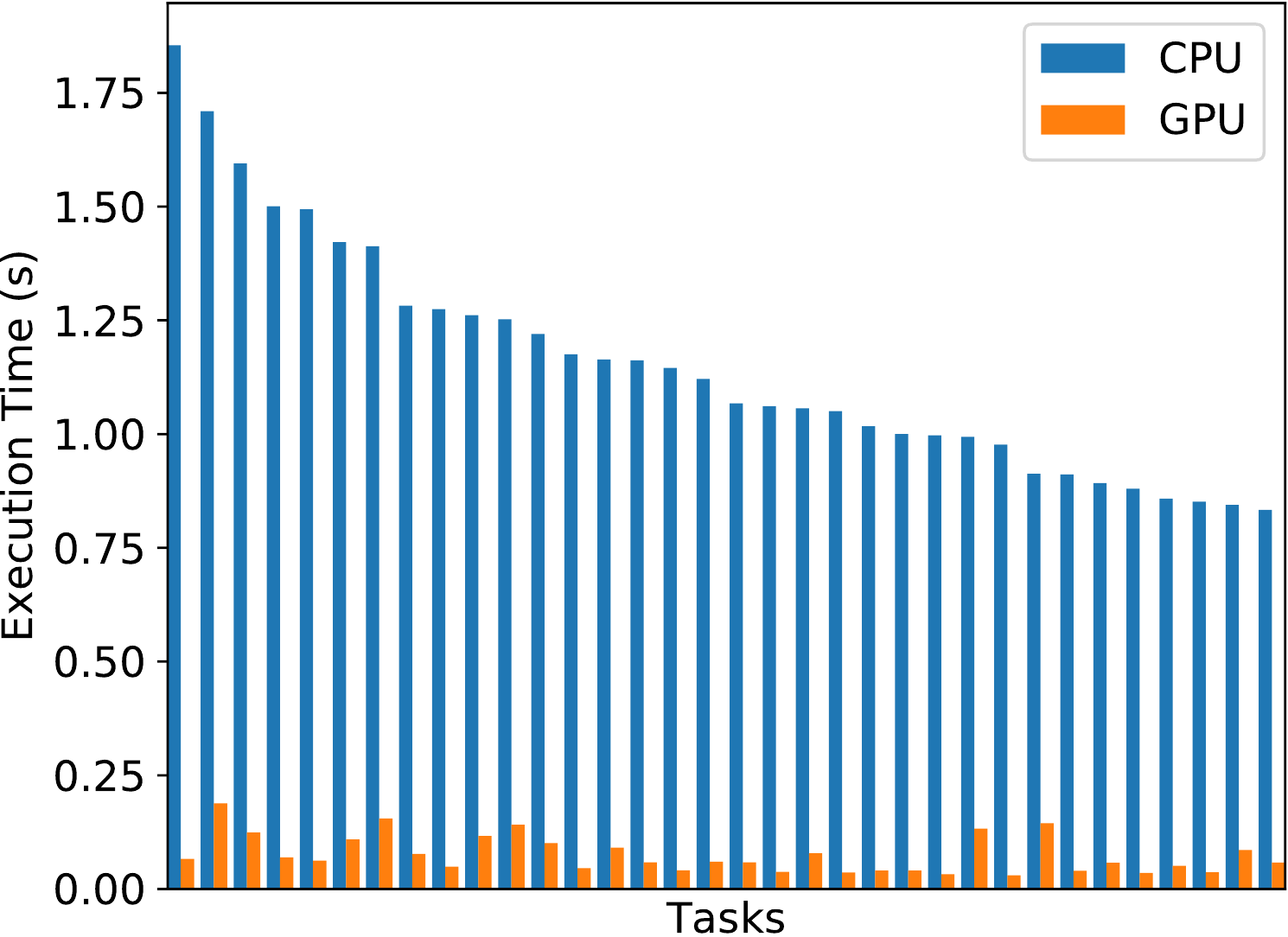}
  \label{fig:taskrun-twi}}

  \subfigure[Friendster; workload estimation]{\includegraphics[width=.45\linewidth]{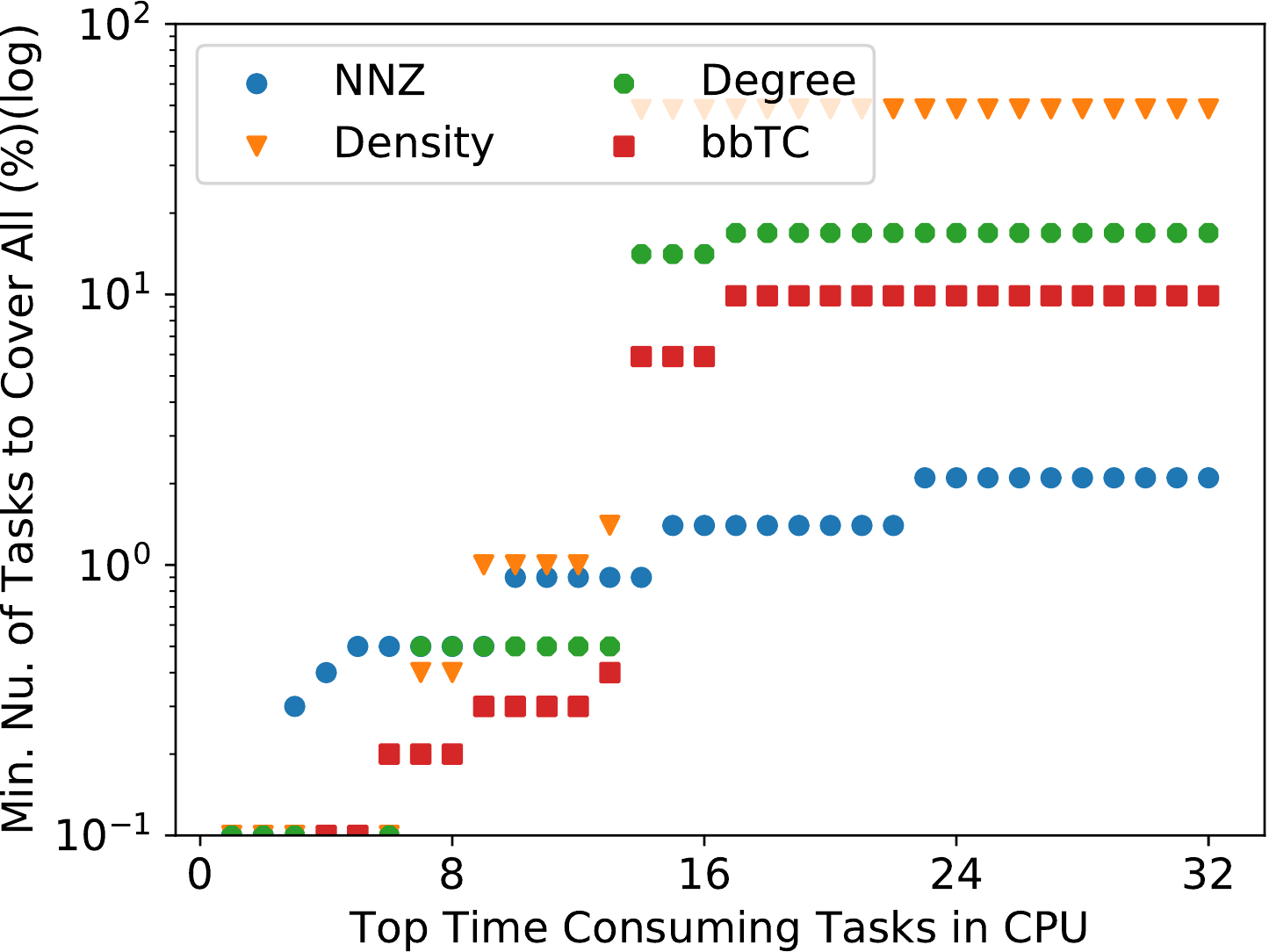}
  \label{fig:estim-fri}}
  \subfigure[Twitter; workload estimation]{\includegraphics[width=.45\linewidth]{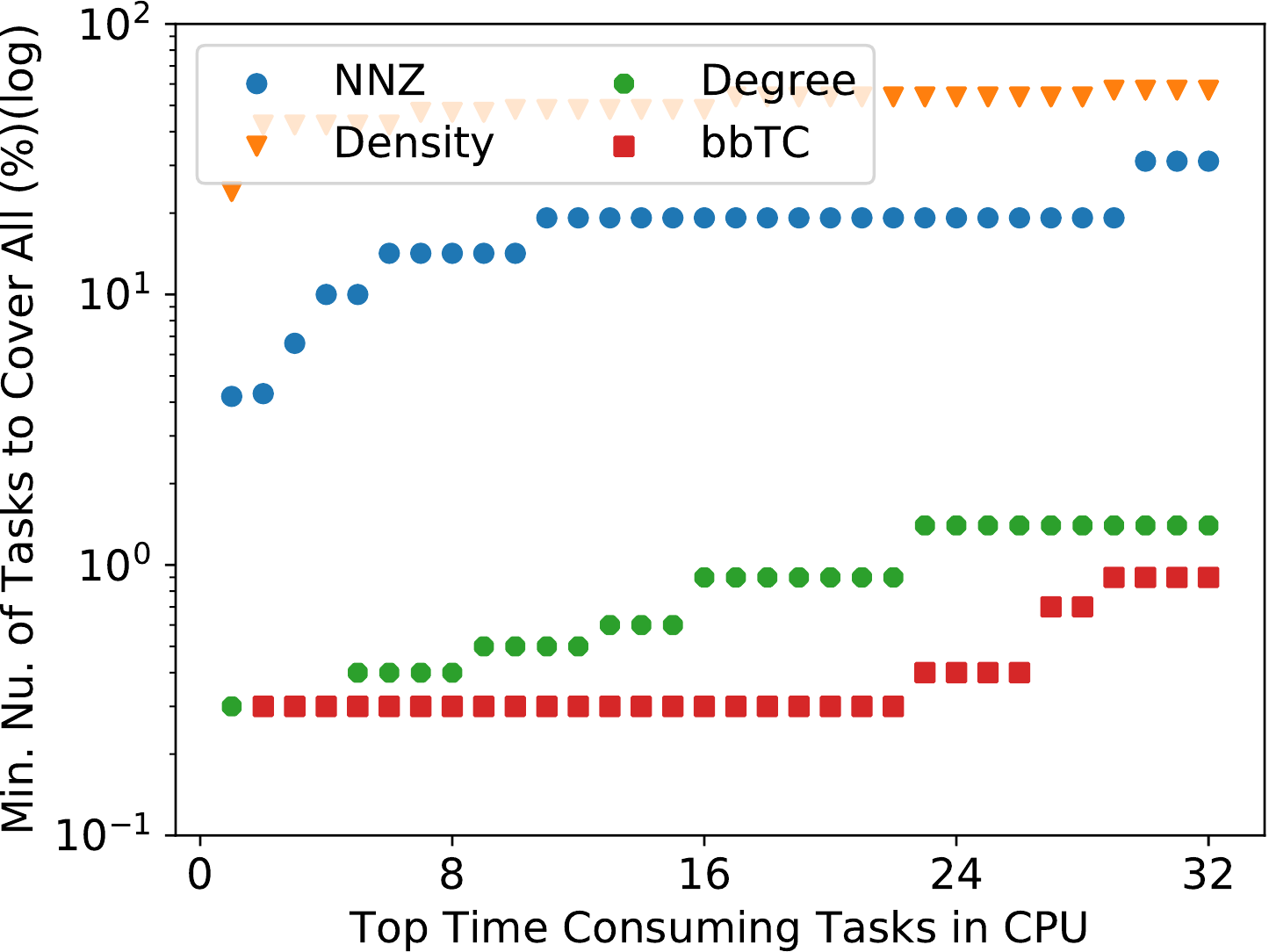}
  \label{fig:estim-twi}}

  \caption{Comparison of Different Estimation Functions}
  \label{fig:esticomp}
\end{figure}

Fig.~\ref{fig:taskrun-fri} and Fig~\ref{fig:taskrun-twi} present execution times
of the most time consuming $32$ tasks (sorted on the x-axis) on a CPU and a GPU for
Friendster and Twitter graphs. We observe that a GPU executes heavy tasks at
least seven times faster than a CPU. Hence, assigning heavy tasks to GPUs is
crucial. Note that, \bbtc doesn't need to estimate workload of tasks with full
accuracy, just identifying heavy tasks is enough to increase GPU utilization.
Therefore, using a thorough model to estimate the workload is an overkill.
In this experiment, we evaluate the effectiveness of our proposed
workload estimation heuristic (see Sec.~\ref{sec:overview}).

Fig.~\ref{fig:estim-fri} and Fig~\ref{fig:estim-twi} present the least required
percentage of tasks (y-axis in log-scale) that starts from the beginning of a
sorted task list using an estimation function, to cover most $x$ time-consuming
tasks (x-axis) (lower is better). In this experiment, we compare four different
estimation functions. NNZ: the total number of nonzeros in a task. Density: the
sum of the average nonzeros per unit in the blocks of a task. Degree: the sum of
the average degree of blocks
in a task and \bbtc's estimation function that we have described in
Sec.~\ref{sec:overview}. We observe that \bbtc's estimation heuristic outperforms
the others in $47$ of $64$ instances. In the Friendster graph, NNZ outperforms
\bbtc's estimation heuristic in predicting the most time consuming $17$ to $32$
tasks. However, on the Twitter graph, NNZ performs poorly. Hence, using NNZ as an
estimation function may end up with unstable performance. In both graphs, not
surprisingly, degree-based estimation performs closer to \bbtc's estimation
heuristic due to the effect of vertex degrees on common neighbor computation.

\subsection{Comparison with TriCore}

\begin{figure}[ht]
  \centering
  \subfigure[Friendster]{\includegraphics[width=.45\linewidth]{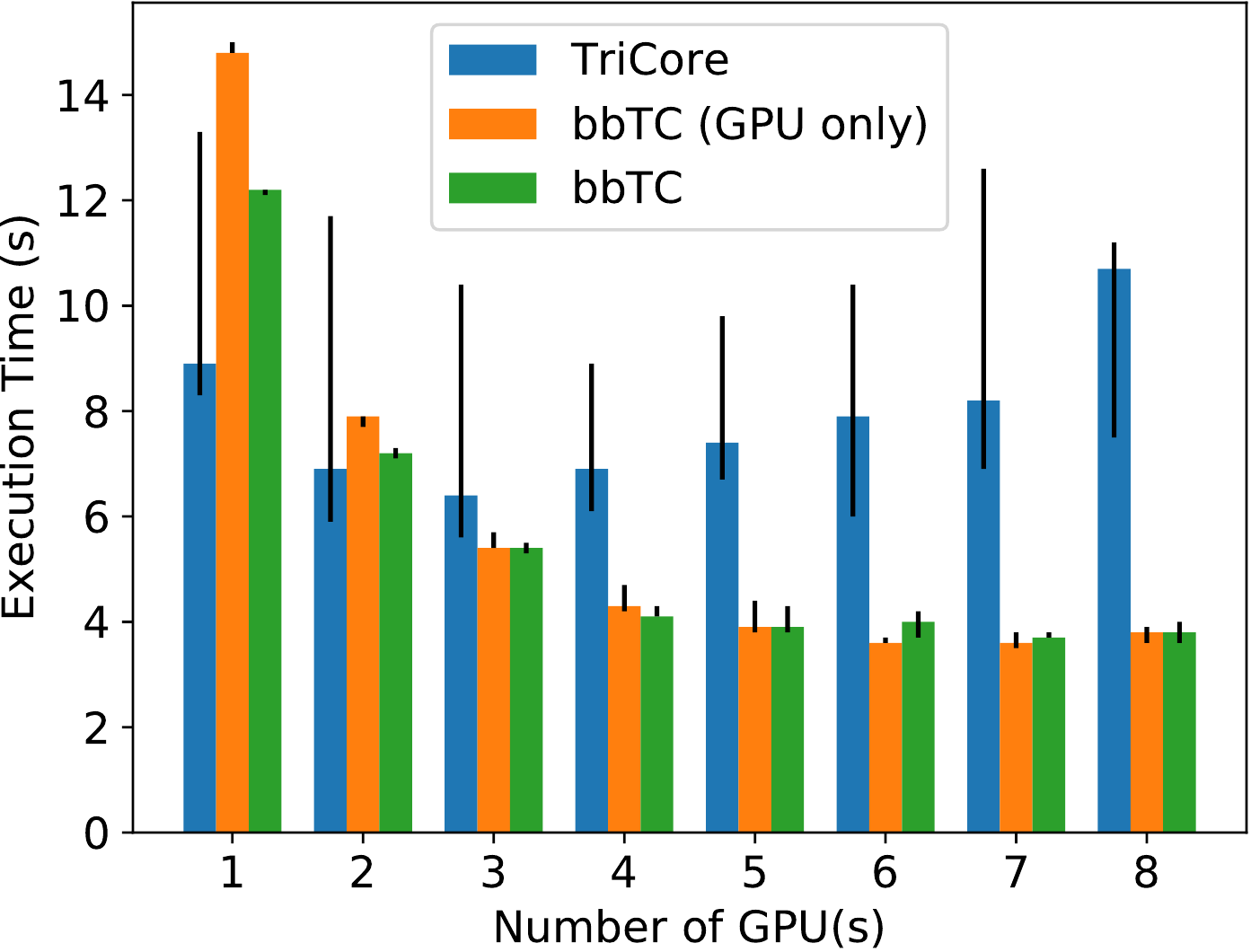}
  \label{fig:tricore-fri}}
  \subfigure[Twitter]{\includegraphics[width=.45\linewidth]{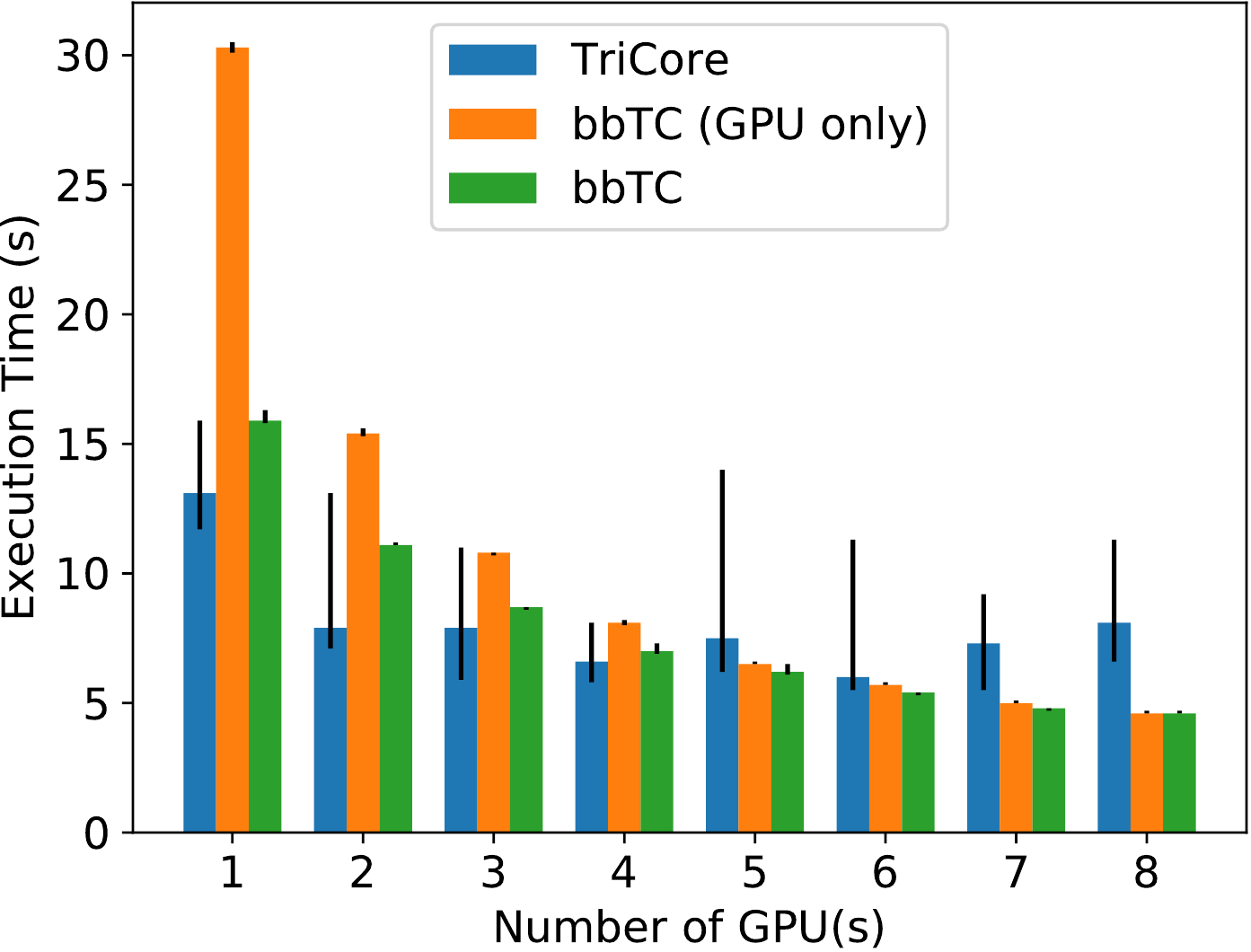}
  \label{fig:tricore-twi}}
  \caption{Comparison with TriCore}
  \label{fig:tricorecomp}
\end{figure}

In this experiment, we compare \bbtc with TriCore using Friendster and Twitter
graphs. Note that, partitioning is turned off for TriCore as it affects the performance
negatively. For Twitter and Friendster graphs we can use TriCore without partitioning
because both graphs can fit into GPUs' memory. However, when one needs to run
larger problem instances, partitioning is going to be necessary and will hurt
TriCore's performance due to bad memory utilization.
Each bar in Fig.~\ref{fig:tricorecomp} illustrates the execution time (y-axis)
of an algorithm for different number of GPUs (x-axis). We observe that \bbtc
scales better than TriCore with respect to the number of GPUs thanks to its
efforts to overlap communication with computation. TriCore's scalability becomes
worse after $4$ GPUs on Friendster, and its times go up after $4$ GPUs on Twitter.
On the other hand, \bbtc scales much better up to $5$ GPUs and continues to
improve slightly, after $5$ GPUs. In Fig.~\ref{fig:tricorecomp}, each error-bar
presents variation between minimum and maximum execution times among five runs.
We observe that the TriCore algorithm is highly unstable, and its runtime deviates
up to $40\%$ while this deviation is less than $5\%$ for \bbtc. TriCore's high
deviation might be caused by unstable warp usage during intersection computation.

Symmetric rectilinear partitioning may generate sparse blocks due to the
irregularity of the graphs and also a generalization of the partitioning. Tasks
that contain highly sparse blocks might be executed extremely fast by CPUs and
GPUs. However, processing these tasks on GPUs, come with an overhead caused by
data copy, API calls, and stream synchronization costs. Hence, assigning all tasks on GPUs
may end up with poor performance and also contradicts with \bbtc's design goals.
\bbtc targets heterogeneous execution. As shown in Fig.~\ref{fig:tricorecomp} because
of the overhead of the light tasks in $1$ and $2$ GPU-only cases, TriCore
outperforms \bbtc. Hybrid execution addresses this issue and \bbtc achieves a competitive performance
in these cases. Note that, even in GPU only case, \bbtc scales better than TriCore
and outperforms it after $4$ GPUs in the system.

\subsection{Overlapping communication and computation}

\begin{figure}[ht]
  \centering
  \subfigure[Friendster]{\includegraphics[width=.45\linewidth]{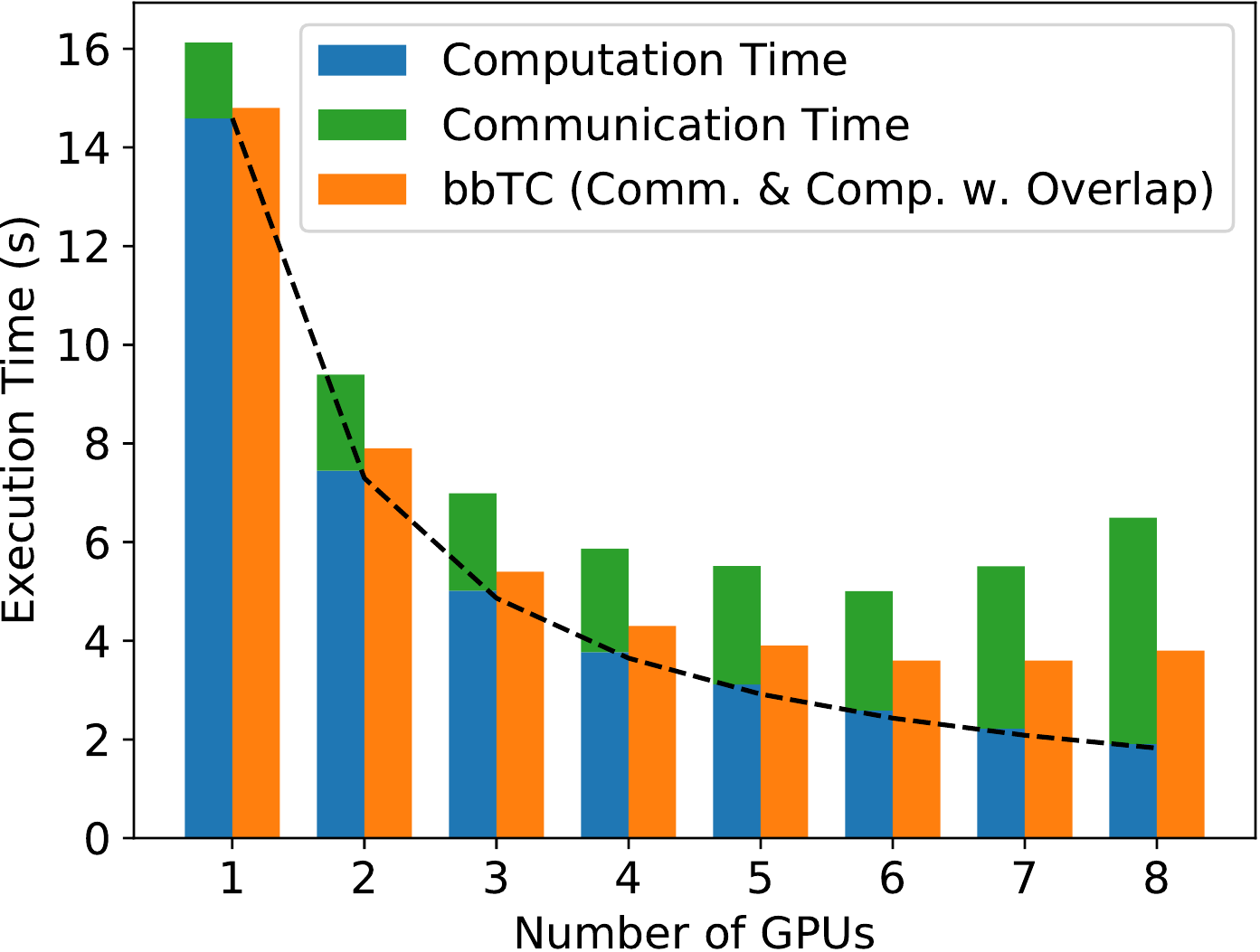}
  \label{fig:overlap-friendster}}
  \hspace{.5em}
  \subfigure[Twitter]{\includegraphics[width=.45\linewidth]{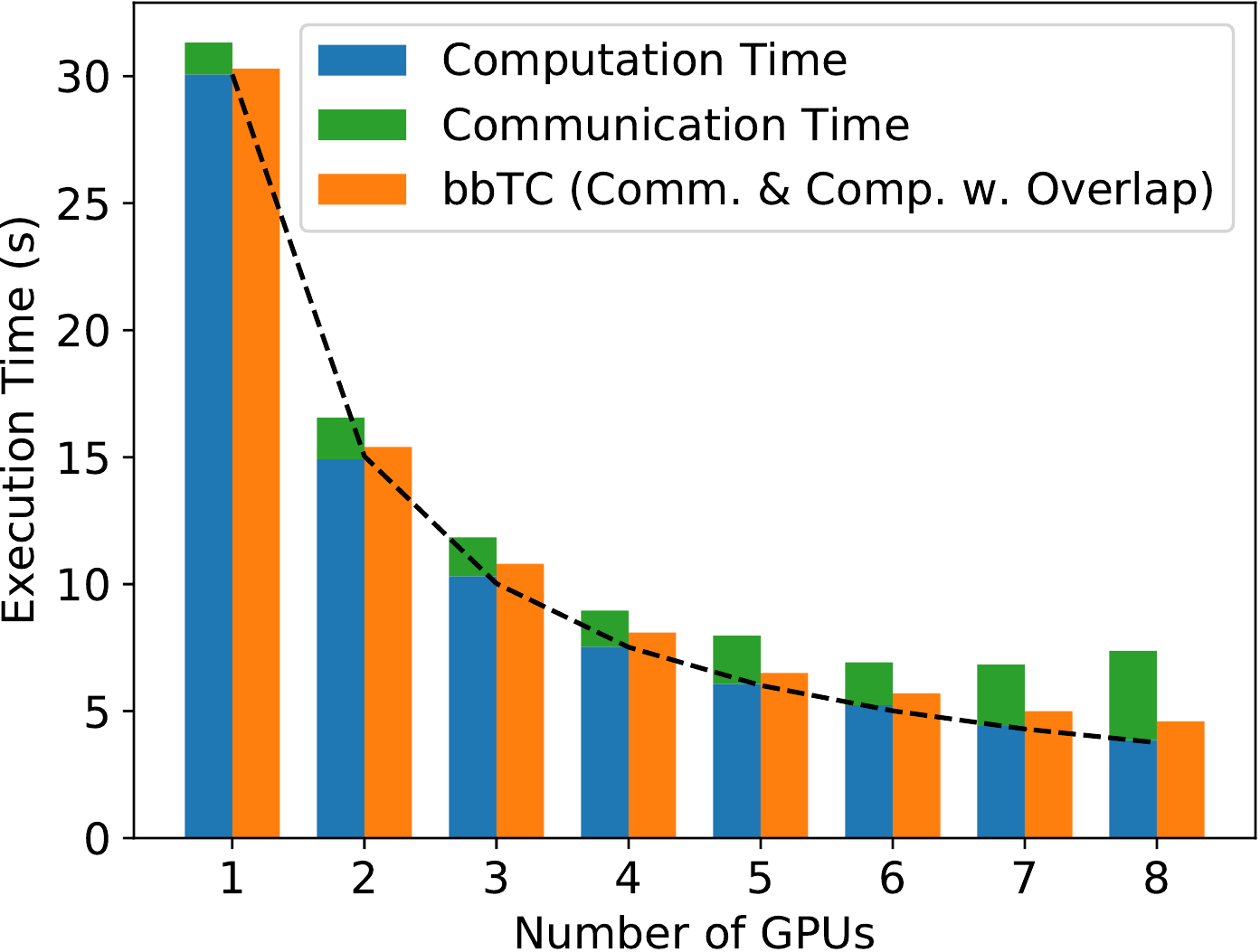}
  \label{fig:overlap-twitter}}
  \caption{Overlap comparison. Dashed line represents linear speedup.}
  \label{fig:overlap}
\end{figure}

Fig.~\ref{fig:overlap} illustrates how successfully \bbtc overlaps communication
with computation in the different number of GPU settings. In
Fig.~\ref{fig:overlap}, blue bars represent computation only execution time
(y-axis), and green bars represent communication time between CPUs and GPUs.
Orange bars represent \bbtc's GPU only execution times, which tries to overlap
communication with computation. The black dashed line represents the linearly
scaled execution time of single GPU computation-only time for the number
of GPUs. As expected, \bbtc's computation time scales almost linearly with
the number of GPUs. We observe that on the Friendster graph, the \bbtc algorithm
can overlap communication with computation with just $10\%$ overhead up to $5$
GPUs, however this overhead increases for the number of GPUs and
becomes $50\%$ when the number of GPUs is $8$, due to bandwidth limitations.
This overlap varies from $1\%$ to $18\%$ on the Twitter graph.

\subsection{Hybrid execution and Cut-off}

\begin{figure}[ht]
  \centering
  \subfigure[DGX (2 GPUs)]{\includegraphics[width=.43\linewidth]{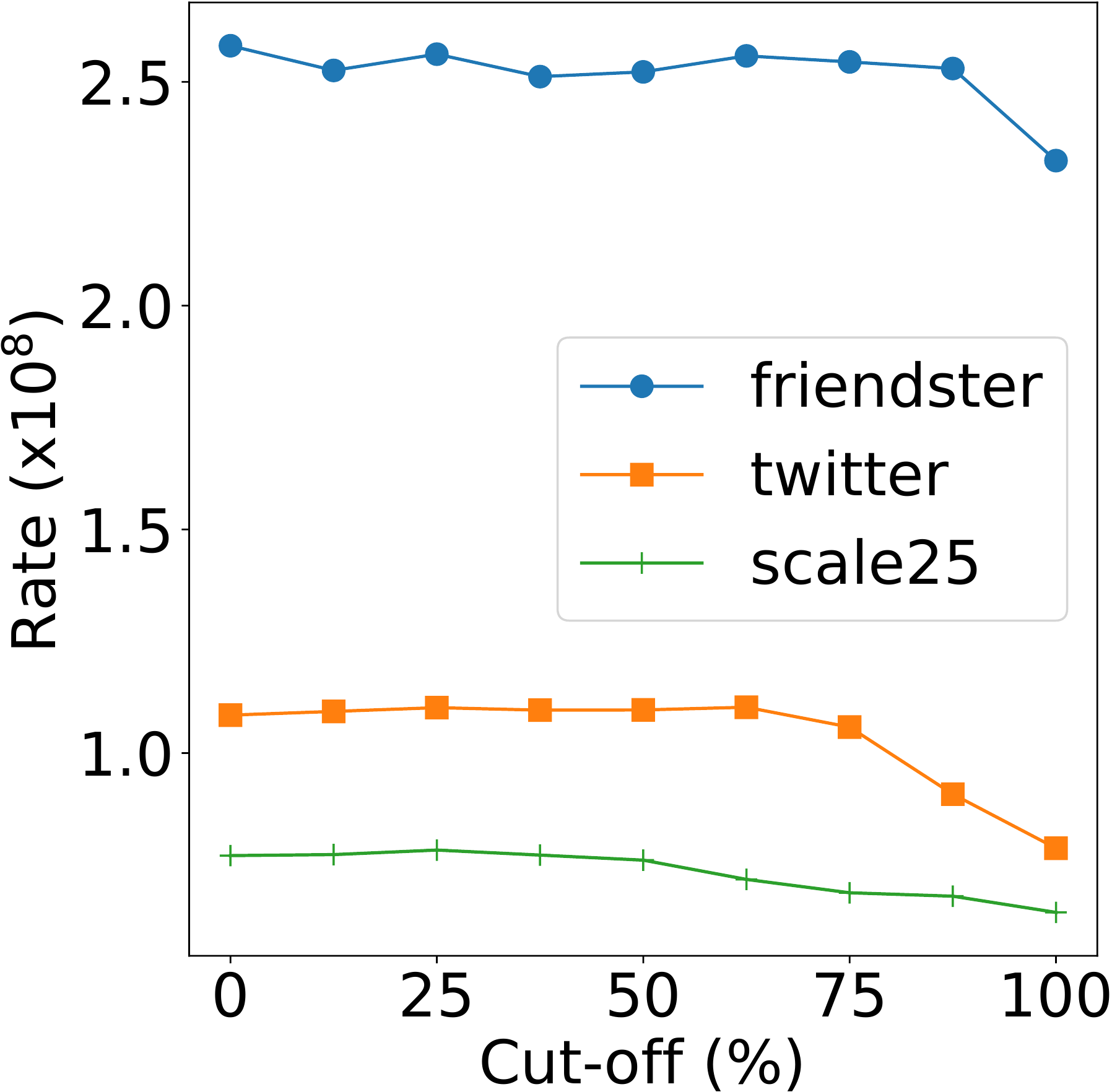}
  \label{fig:dgxborder}}
  \subfigure[Newel (2 GPUs)]{\includegraphics[width=.43\linewidth]{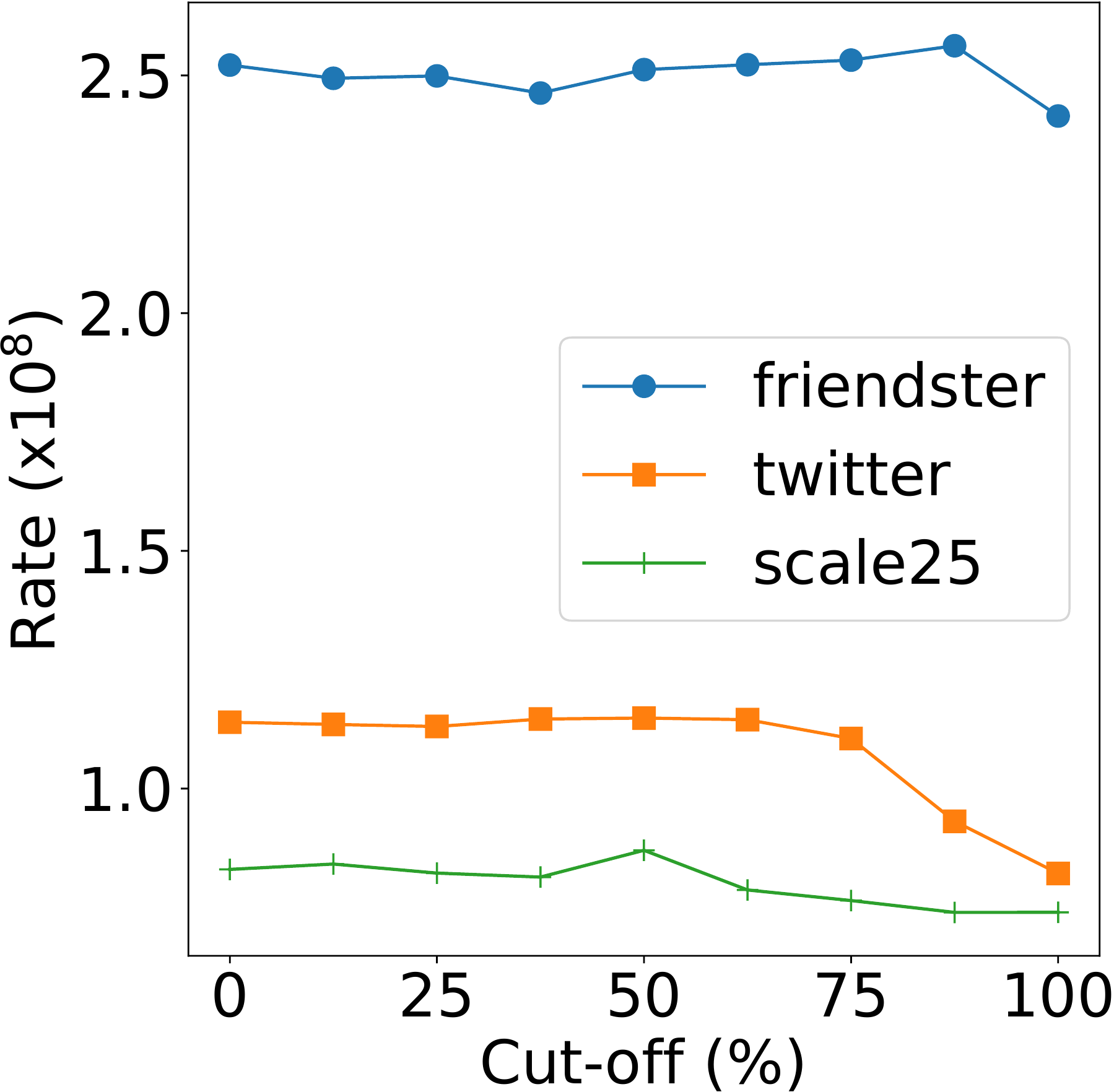}
  \label{fig:newelborder}}

  \caption{Hybrid execution and cut-off.}
  \label{fig:taskeffborder}
\end{figure}

In this experiment, we evaluate the effect of the cut-off location (see Fig.~\ref{fig:hybrid})
from $0$ to $|T|$. If
the cut-off is equal to the number of tasks ($|T|$), we assign all the tasks to
the GPUs. If the cut-off is $0$, then CPUs may execute all the tasks.
In this experiment, we set cut-off as a multiple of one eight of the number
of tasks ($\frac{|T|}{8}$), and report rates for scale25, Twitter, and Friendster
graphs on DGX and Newell architectures. As illustrated in
Fig.~\ref{fig:taskeffborder} when the cut-off is larger than the last two quartiles,
then the overall performance decreases because the execution of the lighter tasks on
GPUs doesn't compensate the overhead.
The cut-off creates a one-way barrier and guarantees the assignment of heavy tasks
on GPU by restricting CPUs for not passing the cut-off point. Hence, the cut-off has
crucial importance. However, thanks to estimation function and relatively fewer
heavy tasks deciding the cut-off point is not a hard problem.
By default, the cut-off point is the middle of the execution queue.

\subsection{Relative speedup}

\begin{figure}[ht]
  \centering
  \includegraphics[width=.9\linewidth]{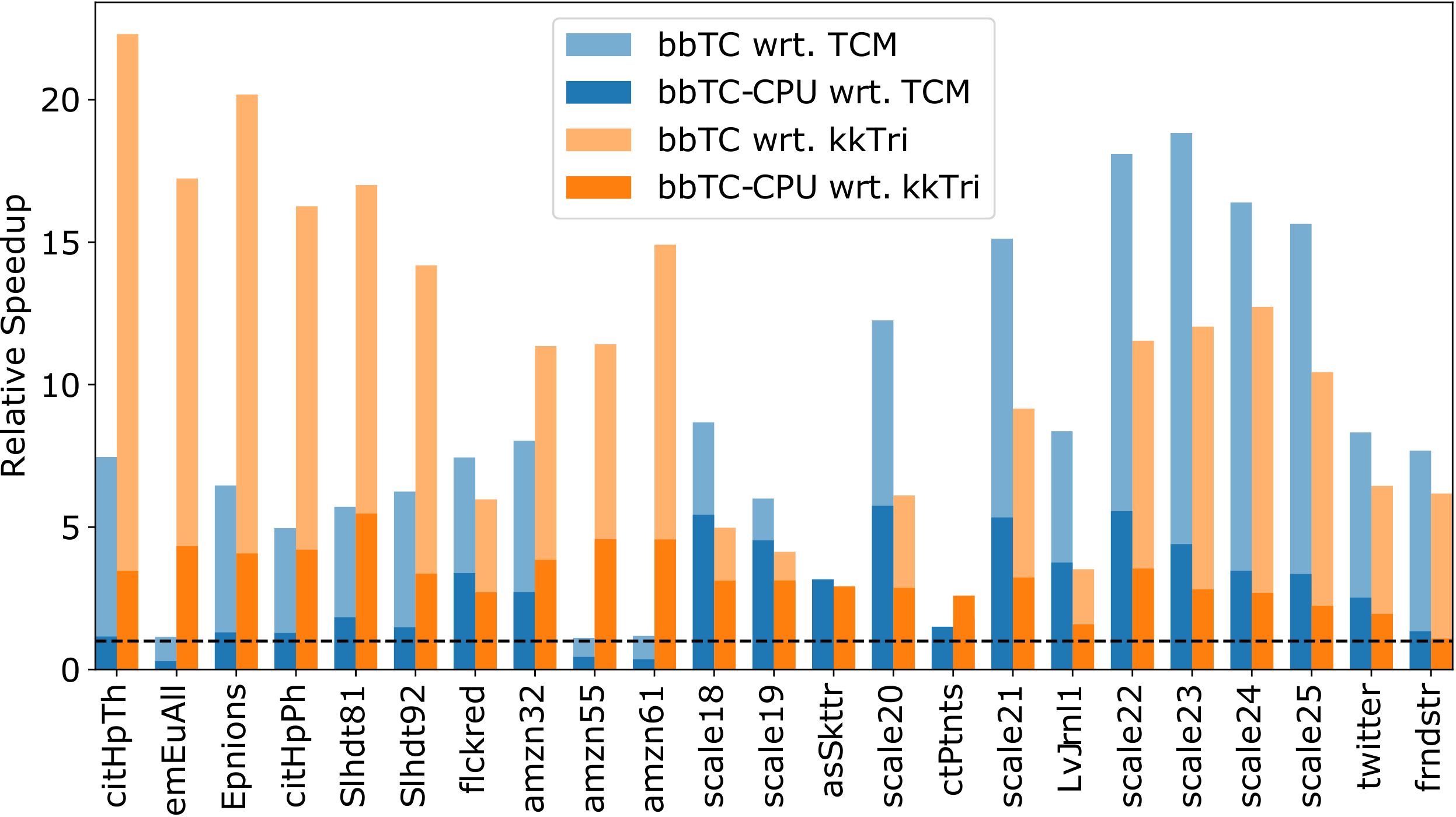}
  \caption{Relative speedup between this work
  and TCM and kkTri algorithms, on Newell architecture. bbTC-CPU: speedup when we only use CPUs.
  bbTC: speedup when we use CPUs and GPUs.
  Black dashed line represents the baseline.
  Graphs are sorted on x-axis based on their number of edges.}
  \label{fig:relative}
\end{figure}

Figure~\ref{fig:relative}  presents  relative  speedup  between this work and
two state-of-the-art algorithms; TCM~\cite{Shun15-ICDE,Dhulipala18-SPAA}, and
kkTri~\cite{Wolf17-HPEC}. We compare \bbtc with these two state-of-the-art
works for CPU only, and heterogeneous scenarios. In the CPU only scenario, \bbtc
outperforms TCM in  $23$ of $23$ cases and kkTri in $20$ of $23$ cases.
kkTri can perform better than \bbtc in three small instances; email-EuAll,
amazon0505 and amazon0611. \bbtc can achieve up to $18\times$ and $14\times$
speedup with respect to TCM and kkTri, respectively.
In the heterogeneous case,
\bbtc outperforms TCM and kkTri in $23$ of $23$ cases.
However, in four small instances (email-EuAll, amazon0505, amazon0611, and
cit-Patents), TCM and \bbtc are comparable. \bbtc can achieve up to $18\times$
speedup on large graphs and up to $22\times$ speedup on smaller instances.

\subsection{Effect of bandwidth on speedup}

\begin{figure}[htb]
  \center
  \includegraphics[width=.55\linewidth]{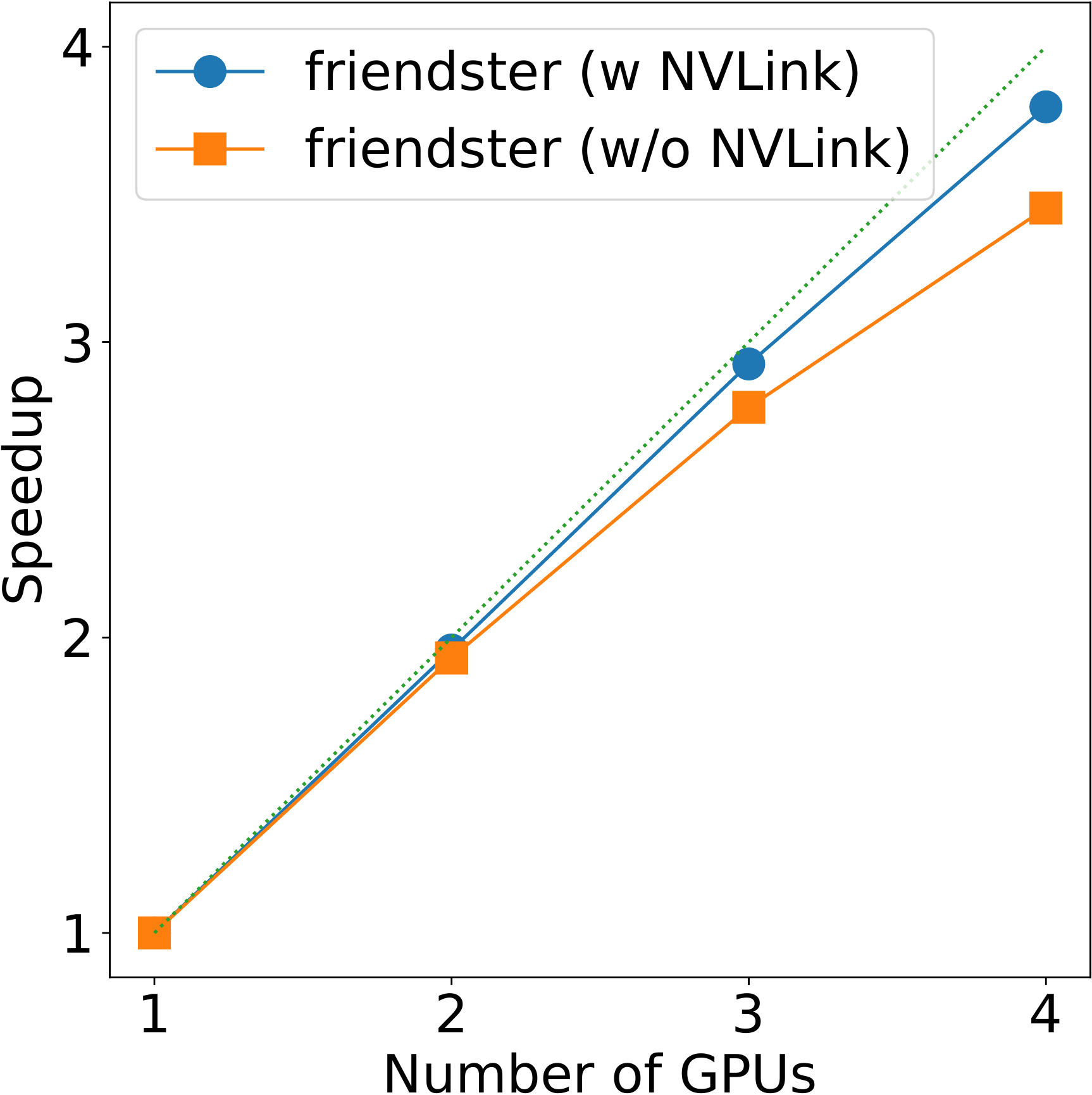}
  \caption{Effect of bandwidth.}
  \label{fig:frsup}
\end{figure}

\bbtc assigns four CUDA streams per GPU. Each of these streams asynchronously
copies the data from the host memory to the device memory. The bandwidth between the host
machine and the device becomes a bottleneck when there are more GPUs on the node.
In this experiment, we evaluate how the bandwidth can affect
the scalability. Fig.~\ref{fig:frsup} illustrates \bbtc's strong scalability
on Friendster graph up to four GPUs on a machine with NVLink and a machine
without NVLink. As expected, with NVLink enabled architecture we get better scaling.
Hence, if we would have a server with NVLink and $8$ Volta GPUs with,
\bbtc's could get even better execution times.

\subsection{Comparison on the WDC-2014 graph}

\begin{table}[htb]
\smallfont
\caption{Execution times of four different algorithms on two different architectures on
a WDC graph.}
\label{table:wdc}

\begin{center}
\begin{tabular}{ l  c | c   }
  & \textbf{Haswell} & \textbf{DGX}\\
  \hline \hline
  \textbf{kkTri} &  423 & Out of memory \\
  \textbf{TCM} &  476 & Out of memory \\
  \textbf{TriCore} &  N/A &  \\
  \textbf{bbTC} &  265 & 116  \\
\end{tabular}
\end{center}
\end{table}

In this experiment, we evaluate the performance of the \bbtc algorithm
on a WDC hyperlink graph. We present runtimes in seconds in Table~\ref{table:wdc}.
This graph has $1.7$ billion vertices and $124$ billion edges. DGX
server doesn't have enough memory to execute this graph. Therefore, kkTri and
TCM algorithms failed on DGX. TriCore also raised an exception during
its partitioning stage; consequently, we couldn't run TriCore. To be able to process
such a graph, which is larger than available memory, \bbtc uses memory-mapped files.
In CPU only execution (Haswell) \bbtc outperforms kkTri and TCM $1.6\times$ and $1.8$
times, respectively. \bbtc can process the WDC-2014 graph in $116$ seconds on
DGX. Note that, on DGX a hardware RAID system is installed. Hence, \bbtc can
perform faster disk I/O.

\section{Conclusions}
\label{sec:conc}
We have developed a triangle counting algorithm called \bbtc,
which leverages a special 2D partitioning scheme called symmetric rectilinear
partitioning to define tasks that can be easily used on any heterogeneous accelerator, CPU combination.
To leverage from the massive computing capabilities of the GPUs and to
overlap communication with computation, \bbtc implements a dynamic task-scheduling
scheme that schedules tasks on multi-core-CPUs and multiple streams on multiple GPUs.
\bbtc is up to $20\times$ faster than state-of-the-art multi-core CPU algorithms and $1.5\times$ faster
than state-of-the-art multi-GPU algorithm. Our experimental results demonstrate that
\bbtc achieves the fastest end-to-end execution times when the graph is not on the GPU.

\vspace{1em}
\noindent {\bf Acknowledgments:}
This work was supported in parts by the NSF grants CCF-1919021 and Sandia
National Laboratories/Sandia Corp contract 2115504.
Sandia National Laboratories is a multi-mission laboratory managed and
operated by National Technology and Engineering Solutions of Sandia,
LLC., a wholly owned subsidiary of Honeywell International, Inc., for
the U.S. Department of Energy's National Nuclear Security Administration
under contract DE-NA-0003525.

\bibliographystyle{IEEEtran}
\bibliography{tdalab,paper}

\begin{thebibliography}{10}
\providecommand{\url}[1]{#1}
\csname url@samestyle\endcsname
\providecommand{\newblock}{\relax}
\providecommand{\bibinfo}[2]{#2}
\providecommand{\BIBentrySTDinterwordspacing}{\spaceskip=0pt\relax}
\providecommand{\BIBentryALTinterwordstretchfactor}{4}
\providecommand{\BIBentryALTinterwordspacing}{\spaceskip=\fontdimen2\font plus
\BIBentryALTinterwordstretchfactor\fontdimen3\font minus
  \fontdimen4\font\relax}
\providecommand{\BIBforeignlanguage}[2]{{%
\expandafter\ifx\csname l@#1\endcsname\relax
\typeout{** WARNING: IEEEtran.bst: No hyphenation pattern has been}%
\typeout{** loaded for the language `#1'. Using the pattern for}%
\typeout{** the default language instead.}%
\else
\language=\csname l@#1\endcsname
\fi
#2}}
\providecommand{\BIBdecl}{\relax}
\BIBdecl

\bibitem{Latapy08-TCS}
M.~Latapy, ``Main-memory triangle computations for very large (sparse
  (power-law)) graphs,'' \emph{Theoretical Computer Science}, pp. 458--473,
  2008.

\bibitem{Shun15-ICDE}
J.~Shun and K.~Tangwongsan, ``Multicore triangle computations without tuning,''
  in \emph{2015 IEEE 31st International Conference on Data Engineering}.\hskip
  1em plus 0.5em minus 0.4em\relax IEEE, 2015, pp. 149--160.

\bibitem{Hu18-SC}
Y.~Hu, H.~Liu, and H.~H. Huang, ``Tricore: Parallel triangle counting on
  gpus,'' in \emph{SC18: International Conference for High Performance
  Computing, Networking, Storage and Analysis}.\hskip 1em plus 0.5em minus
  0.4em\relax IEEE, 2018, pp. 171--182.

\bibitem{Dhulipala18-SPAA}
L.~Dhulipala, G.~E. Blelloch, and J.~Shun, ``Theoretically efficient parallel
  graph algorithms can be fast and scalable,'' in \emph{Proceedings of the 30th
  on Symposium on Parallelism in Algorithms and Architectures}.\hskip 1em plus
  0.5em minus 0.4em\relax ACM, 2018, pp. 393--404.

\bibitem{Berry14-TCS}
J.~W. Berry, L.~K. Fostvedt, D.~J. Nordman, C.~A. Phillips, C.~Seshadhri, and
  A.~G. Wilson, ``Why do simple algorithms for triangle enumeration work in the
  real world?'' in \emph{Theoretical Computer Science}, 2014, pp. 225--234.

\bibitem{Itai78-SJC}
A.~Itai and M.~Rodeh, ``Finding a minimum circuit in a graph,'' \emph{SIAM
  Journal on Computing}, vol.~7, no.~4, pp. 413--423, 1978.

\bibitem{Ortmann14-ALENEX}
M.~Ortmann and U.~Brandes, ``Triangle listing algorithms: Back from the
  diversion,'' in \emph{Proceedings of the Meeting on Algorithm Engineering \&
  Expermiments}.\hskip 1em plus 0.5em minus 0.4em\relax SIAM, 2014, pp. 1--8.

\bibitem{Alon97-Algorithmica}
N.~Alon, R.~Yuster, and U.~Zwick, ``Finding and counting given length cycles,''
  \emph{Algorithmica}, vol.~17, no.~3, pp. 209--223, 1997.

\bibitem{Pagh14-PODS}
R.~Pagh and F.~Silvestri, ``The input/output complexity of triangle
  enumeration,'' in \emph{Proceedings of the 33rd ACM SIGMOD-SIGACT-SIGART
  symposium on Principles of database systems}.\hskip 1em plus 0.5em minus
  0.4em\relax ACM, 2014, pp. 224--233.

\bibitem{Cohen08-NSATR}
J.~Cohen, ``Trusses: Cohesive subgraphs for social network analysis,''
  \emph{National security agency technical report}, pp. 3--1, 2008.

\bibitem{Prat12-CIKM}
A.~Prat-P{\'e}rez, D.~Dominguez-Sal, J.~M. Brunat, and J.-L. Larriba-Pey,
  ``Shaping communities out of triangles,'' in \emph{Proceedings of the 21st
  ACM international conference on Information and knowledge management}.\hskip
  1em plus 0.5em minus 0.4em\relax ACM, 2012, pp. 1677--1681.

\bibitem{Eckmann02-NAS}
J.-P. Eckmann and E.~Moses, ``Curvature of co-links uncovers hidden thematic
  layers in the world wide web,'' \emph{Proceedings of the national academy of
  sciences}, vol.~99, no.~9, pp. 5825--5829, 2002.

\bibitem{Wang10-VLDBJ}
N.~Wang, J.~Zhang, K.-L. Tan, and A.~K. Tung, ``On triangulation-based dense
  neighborhood graph discovery,'' \emph{Proceedings of the VLDB Endowment}, pp.
  58--68, 2010.

\bibitem{Tsourakakis11-SNAN}
C.~E. Tsourakakis, P.~Drineas, E.~Michelakis, I.~Koutis, and C.~Faloutsos,
  ``Spectral counting of triangles via element-wise sparsification and
  triangle-based link recommendation,'' \emph{Social Network Analysis and
  Mining}, pp. 75--81, 2011.

\bibitem{Date17-HPEC}
K.~Date, K.~Feng, R.~Nagi, J.~Xiong, N.~S. Kim, and W.-M. Hwu, ``Collaborative
  ({CPU} + {GPU}) algorithms for triangle counting and truss decomposition on
  the minsky architecture: Static graph challenge: Subgraph isomorphism,'' in
  \emph{2017 IEEE High Performance Extreme Computing Conference (HPEC)}.\hskip
  1em plus 0.5em minus 0.4em\relax IEEE, 2017, pp. 1--7.

\bibitem{Yasar19-HPEC}
A.~Ya{\c{s}}ar, S.~Rajamanickam, J.~W. Berry, M.~M. Wolf, J.~Young, and
  {\"U}.~V. {\c{C}}ataly{\"u}rek, ``Linear algebra-based triangle counting via
  fine-grained tasking on heterogeneous environments,'' in \emph{High
  Performance Extreme Computing Conference (HPEC), 2019 IEEE}.\hskip 1em plus
  0.5em minus 0.4em\relax IEEE, 2019.

\bibitem{Green14-WIAP}
O.~Green, P.~Yalamanchili, and L.-M. Mungu{\'\i}a, ``Fast triangle counting on
  the {GPU},'' in \emph{Proceedings of the 4th Workshop on Irregular
  Applications: Architectures and Algorithms}.\hskip 1em plus 0.5em minus
  0.4em\relax IEEE Press, 2014, pp. 1--8.

\bibitem{Yasar18-HPEC}
\BIBentryALTinterwordspacing
A.~Ya{\c{s}}ar, S.~Rajamanickam, M.~M. Wolf, J.~W. Berry, and {\"U}.~V.
  {\c{C}}ataly{\"u}rek, ``Fast triangle counting using cilk,'' in \emph{High
  Performance Extreme Computing Conference (HPEC), 2018 IEEE}.\hskip 1em plus
  0.5em minus 0.4em\relax IEEE, 2018, pp. 1--7. [Online]. Available:
  \url{http://ieeexplore.ieee.org/document/8547563}
\BIBentrySTDinterwordspacing

\bibitem{Boman13-SC}
E.~G. Boman, K.~D. Devine, and S.~Rajamanickam, ``Scalable matrix computations
  on large scale-free graphs using 2d graph partitioning,'' in \emph{SC'13:
  Proceedings of the International Conference on High Performance Computing,
  Networking, Storage and Analysis}, 2013, pp. 1--12.

\bibitem{Saule12-JPDC-spart}
\BIBentryALTinterwordspacing
E.~Saule, E.~O. Bas, and {\"{U}}.~V. {\c{C}}ataly{\"{u}}rek, ``Load-balancing
  spatially located computations using rectangular partitions,'' \emph{Journal
  of Parallel and Distributed Computing}, vol.~72, no.~10, pp. 1201--1214,
  2012. [Online]. Available: \url{http://dx.doi.org/10.1016/j.jpdc.2012.05.013}
\BIBentrySTDinterwordspacing

\bibitem{Catalyurek10-SISC}
\BIBentryALTinterwordspacing
{\"{U}}.~V. {\c{C}}ataly{\"{u}}rek, C.~Aykanat, and B.~U{\c{c}}ar, ``On
  two-dimensional sparse matrix partitioning: Models, methods, and a recipe,''
  \emph{{SIAM} Journal on Scientific Computing (SISC)}, vol.~32, no.~2, pp.
  656--683, 2010. [Online]. Available:
  \url{http://dx.doi.org/10.1137/080737770}
\BIBentrySTDinterwordspacing

\bibitem{kayaaslan20181}
E.~Kayaaslan, C.~Aykanat, and B.~U{\c{c}}ar, ``{1.5D} parallel sparse
  matrix-vector multiply,'' \emph{SIAM Journal on Scientific Computing},
  vol.~40, no.~1, pp. C25--C46, 2018.

\bibitem{acer2018optimizing}
S.~Acer, O.~Selvitopi, and C.~Aykanat, ``Optimizing nonzero-based sparse matrix
  partitioning models via reducing latency,'' \emph{Journal of Parallel and
  Distributed Computing}, vol. 122, pp. 145--158, 2018.

\bibitem{Grigni96-PAISP}
M.~Grigni and F.~Manne, ``On the complexity of the generalized block
  distribution,'' in \emph{International Workshop on Parallel Algorithms for
  Irregularly Structured Problems}.\hskip 1em plus 0.5em minus 0.4em\relax
  Springer, 1996, pp. 319--326.

\bibitem{Yasar19-arXiv}
\BIBentryALTinterwordspacing
A.~Ya{\c{s}}ar and {\"{U}}.~V. {\c{C}}ataly{\"{u}}rek, ``Heuristics for
  symmetric rectilinear matrix partitioning,'' ArXiv, Tech. Rep.
  arXiv:1909.12209, Oct 2019. [Online]. Available:
  \url{http://arxiv.org/abs/1909.12209}
\BIBentrySTDinterwordspacing

\bibitem{Oleary85COMM}
D.~P. O'leary and G.~Stewart, ``Data-flow algorithms for parallel matrix
  computation,'' \emph{Communications of the ACM}, vol.~28, no.~8, pp.
  840--853, 1985.

\bibitem{Hendrickson950IJHSC}
B.~Hendrickson, R.~Leland, and S.~Plimpton, ``An efficient parallel algorithm
  for matrix-vector multiplication,'' \emph{International Journal of High Speed
  Computing}, vol.~7, no.~01, pp. 73--88, 1995.

\bibitem{Im04-JHPCA}
E.-J. Im, K.~Yelick, and R.~Vuduc, ``Sparsity: Optimization framework for
  sparse matrix kernels,'' \emph{The International Journal of High Performance
  Computing Applications}, vol.~18, no.~1, pp. 135--158, 2004.

\bibitem{Teodoro10-HPDC}
\BIBentryALTinterwordspacing
G.~Teodoro, T.~D.~R. Hartley, {\"{U}}.~V. {\c{C}}ataly{\"{u}}rek, and
  R.~Ferreira, ``Run-time optimizations for replicated dataflows on
  heterogeneous environments,'' in \emph{Proc. of the 19th ACM International
  Symposium on High Performance Distributed Computing (HPDC)}, 2010, pp.
  13--24. [Online]. Available: \url{http://portal.acm.org/Graph Based Citation
  Analysis.cfm?id=1851479}
\BIBentrySTDinterwordspacing

\bibitem{Wolf17-HPEC}
M.~M. Wolf, M.~Deveci, J.~W. Berry, S.~D. Hammond, and S.~Rajamanickam, ``{Fast
  linear algebra-based triangle counting with kokkoskernels},'' in \emph{IEEE
  High Performance extreme Computing Conference (HPEC)}.\hskip 1em plus 0.5em
  minus 0.4em\relax IEEE, 2017, pp. 1--7.

\bibitem{Deveci17-IPDPSW}
M.~Deveci, C.~Trott, and S.~Rajamanickam, ``Performance-portable sparse
  matrix-matrix multiplication for many-core architectures,'' in \emph{Parallel
  and Distributed Processing Symposium Workshops (IPDPSW)}.\hskip 1em plus
  0.5em minus 0.4em\relax IEEE, 2017, pp. 693--702.

\bibitem{Tom19-ICPP}
A.~S. Tom and G.~Karypis, ``A {2D} parallel triangle counting algorithm for
  distributed-memory architectures,'' in \emph{International Conference on
  Parallel Processing}.\hskip 1em plus 0.5em minus 0.4em\relax ACM, 2019,
  p.~45.

\bibitem{Acer19-HPEC}
S.~Acer, A.~Ya{\c{s}}ar, S.~Rajamanickam, M.~M. Wolf, and {\"U}.~V.
  {\c{C}}ataly{\"u}rek, ``Scalable triangle counting on distributed-memory
  systems,'' in \emph{High Performance Extreme Computing Conference (HPEC),
  2019 IEEE}.\hskip 1em plus 0.5em minus 0.4em\relax IEEE, 2019.

\bibitem{Hu18-HPEC}
Y.~Hu, H.~Liu, and H.~H. Huang, ``High-performance triangle counting on gpus,''
  in \emph{IEEE High Performance extreme Computing Conference (HPEC)}.\hskip
  1em plus 0.5em minus 0.4em\relax IEEE, 2018, pp. 1--5.

\bibitem{Davis11-TOMS}
T.~A. Davis and Y.~Hu, ``{The University of Florida sparse matrix
  collection},'' \emph{ACM Transactions on Mathematical Software (TOMS)}, p.~1,
  2011.

\bibitem{Dolan02-MP}
E.~D. Dolan and J.~J. Mor{\'e}, ``Benchmarking optimization software with
  performance profiles,'' \emph{Mathematical programming}, vol.~91, no.~2, pp.
  201--213, 2002.

\end{thebibliography}

\begin{IEEEbiographynophoto}{Abdurrahman Ya\c{s}ar}
is a PhD Student in the School of Computational Science and Engineering at
Georgia Institute of Technology. He received his M.S. in Computer Engineering
from Bilkent University, Turkey in 2015.
\end{IEEEbiographynophoto}

\begin{IEEEbiographynophoto}{Sivasankaran Rajamanickam}
(M’14) is a Principal Member of Technical Staff in the Center for Computing
Research at Sandia National Laboratories. He earned his B.E. from Madurai
Kamaraj University, India, and his Ph.D. in Computer Engineering from University
of Florida.
\end{IEEEbiographynophoto}

\begin{IEEEbiographynophoto}{Jonathan Berry}
is a Distinguished Member of the Technical Staff at Sandia National
Laboratories.  He holds a Ph.D. in computer science from Rensselaer Polytechnic
Institute and spent almost a decade in liberal arts academia before joining
Sandia in 2004.
\end{IEEEbiographynophoto}

\begin{IEEEbiographynophoto}{\"Umit V. \c{C}ataly\"urek}
(M’09,SM’10,Fellow’16) is a Professor and Associate Chair in the School of
Computational Science and Engineering at the Georgia Institute of Technology. He
received his Ph.D., M.S. and B.S. in Computer Engineering and Information
Science from Bilkent University, Turkey.
\end{IEEEbiographynophoto}

\end{document}